\documentclass[iop,preprint,10pt]{emulateapj}
\usepackage[]{natbib}
\usepackage{graphicx}
\usepackage{subfigure}
\usepackage[dvipsnames]{color}
\usepackage{float}
\usepackage{amsmath}
\usepackage{fancyref}
\definecolor{orange}{RGB}{255,69,0}
\definecolor{green}{RGB}{0,255,0}
\definecolor{darkred}{RGB}{139,0,0}
\usepackage{amssymb}
\usepackage{lineno}
\usepackage[colorlinks=true,
            linkcolor=orange,
            urlcolor=magenta,
            citecolor=blue]{hyperref}
\usepackage{nameref}
\usepackage{array,multirow,makecell}
\usepackage{tabularx}
\usepackage{booktabs}

\usepackage{morefloats}
\usepackage{gensymb}
\begin{document}
\title
{Broadband variability and correlation study of 3C 279 during flare of 2017-2018  }

\author{Raj Prince$^{1}$}
\affil{$^{1}$Raman Research Institute, Sadashivanagar, Bangalore 560080, India \\
        } 

\email{rajprince@rri.res.in}
\begin{abstract}
A multiwavelength temporal and spectral analysis of flares of 3C 279 during November 2017--July 2018 are presented in this work. Three bright 
gamma-ray flares were
observed simultaneously in X-ray and Optical/UV along with a prolonged quiescent state. A ``harder-when-brighter'' trend is observed in both gamma-rays and
X-rays during the flaring period. The gamma-ray light curve for all the flares are binned in one-day time bins and a day scale variability is observed.
Variability time constrains the size and location of the emission region to 2.1$\times$10$^{16}$ cm and 4.4$\times$10$^{17}$ cm, respectively.
The fractional variability reveals that the source is more than 100\% variable in gamma-rays and it decreases towards the lower energy. 
A cross-correlation study of the emission from different wavebands is done using the \textit{DCF} method, which shows a strong correlation
between them without any time lags. The zero time 
lag between different wavebands suggest their co-spatial origin. This is the first time 3C 279 has shown a strong correlation between gamma-rays 
and X-rays emission with zero time lag. A single zone emission model was adopted to model the multiwavelength SEDs by using
the publicly available code GAMERA. 
The study reveals that a higher jet power in electrons is required to explain the gamma-ray flux during the flaring state, as much as, ten times of 
that required for the quiescent state.
However, more jet power in magnetic field has been observed during the quiescent state compared to the flaring state. 

\end{abstract}
%\keywords{galaxies: active; gamma rays: galaxies; individuals: 3C 279}
\keywords{galaxies: active -- galaxies: jets -- gamma rays: galaxies -- quasars: individual (3C 279)}

\section{Introduction}
Blazars are a class of active galactic nuclei whose jets are oriented close to the observer's line of sight \citep{Urry and Padovani (1995)}. 
In general, blazars exhibit highly luminous and rapidly variable non-thermal continuum emission across the entire electromagnetic spectrum 
extending from radio to very high energy gamma-ray. 
A wide range of variability time across the whole electromagnetic spectrum is found in most of the blazars. 
A time scale of variability ranging from minutes to years is inferred for blazars (\citealt{Aharonian et al. (2007)}; \citealt{Raiteri et al. (2013)}). 
Blazars are generally classified into two categories viz., BL Lac objects (BL Lacs) and flat spectrum radio quasar FSRQ, depending upon their optical spectra. 
BL Lacs shows a very weak or no emission line in their optical spectra while on the other hand FSRQs are known for their strong, broad
emission lines. The highly energetic phenomenon inside the blazars are detected as strong and spectacular flares across the entire electromagnetic
spectrum, with rapid variability.
There have been several studies on blazar to understand the broadband flaring activity, but the origin of fast variability is not well understood.
The observed broadband spectral energy distribution (SED) of blazar shows two peaks, one extending from radio to optical, and another
ranging from X-ray to gamma-rays. The low energy radiation from radio to UV/X-ray is caused by synchrotron radiation of relativistic electrons
accelerated inside the jet. In leptonic models, the high energy gamma-ray radiation is caused by inverse Compton (IC) scattering of soft target 
photons originating in synchrotron radiation (SSC; \citealt{Sikora et al. (2009)}) or external photon fields (EC; \citealt{Dermer et al. (1992)}; \citealt{Sikora et al. (1994)}).

3C 279 has been classified as a FSRQ at redshift, $z$ of 0.536 \citep{Lynds et al. (1965)}, and is a well-studied blazar in its class. 
The mass of the black hole was estimated in the range of (3--8)$\times$10$^{8}$M$_\odot$ by \citet{Woo and Urry (2002)}, \citet{Gu et al. (2001)},
and \citet{Nilsson et al. (2009)}
from the luminosity of broad optical emission lines, the width of the H$_\beta$ lines, and from the luminosity of the host galaxy respectively.
Fermi-LAT is continuously monitoring 3C 279 since 2008 and along with that it is also monitored by different other facilities in X-ray, optical, and
radio. 
The blazar 3C 279 has been studied in great extent in past throughout the entire electromagnetic spectrum. The previous multi-wavelength
studies shows that the gamma-ray emission region in 3C 279 lies close to the base of the jet, at sub-parsec scale (\citealt{Hayashida et al. (2015)}; 
\citealt{Paliya (2015a)}; \citealt{Paliya et al. (2016)}). However, the recent study by \citet{Patino-Alvarez et al. (2019)} found an evidence of
high energy emission emitted at much larger distance from the core, at $\sim$ 42 pc. Further, multiwavelength study on this source will help the
community to probe the location of the gamma-ray emission region along the jet axis.

The correlation study between optical polarization (degree/angle) and gamma-rays give strong evidence for the synchrotron and Compton
model, and suggest that the jet structure is not axisymmetric \citep{Abdo et al. (2010)}. The above correlation also suggest the presence of 
helical magnetic field component \citep{Zhang et al. (2015)}. 
The rate of change of the polarization angle with fast variability time suggests a compact emission region, close to the central black hole, 
located at a distance along the jet. A detailed correlations study between optical and gamma-ray has been done for a sample of blazars by 
\citet{Cohen et al. (2014)}, and they have found a time lag of 110 days between optical and gamma-ray emission. A similar study has been done by
\citet{Hayashida et al. (2012)} for 3C 279 and they have noticed a lag of 10 days between optical and gamma-ray emission. They also found that the 
X-ray and gamma-ray emissions are not well correlated, and the nature of X-ray emission in blazar is still unclear.
A correlation study between radio and gamma-rays is also done by \citet{Pushkarev et al. (2010)}, and they noticed that in a sample of
183 blazars most of them show that the radio flares lag the gamma-ray flare. A similar result was found in case of Ton 599 by 
\citet{Prince (2019)}, where a lag of 27 days was noticed between radio and gamma-rays. Keeping consistency with results from 
earlier studies, here I try to investigate the possible correlation between different wavelength during the 2017-2018 flare of 3C 279.\\

\section{Multiwavelength Observations and Data Analysis}
\subsection{\textit{Fermi}-LAT}
\textit{Fermi}-LAT is a pair conversion $\gamma$-ray Telescope sensitive to photon energies between 20 MeV
to higher than 500 GeV, with a field of view of about 2.4 sr \citep{Atwood et al. (2009)}. The LAT's field of view covers about
20\% of the sky at any time, and it scans the whole sky every three hours.
NASA launched the instrument in 2008 into a near earth orbit. 
\textit{Fermi}-LAT is continuously monitoring 3C 279 since 2008 August. 
The standard data reduction and analysis procedure\footnote{https://fermi.gsfc.nasa.gov/ssc/data/analysis/documentation/} 
has been followed. I have chosen a circular region of radius 10$\degree$ around the source of interest during the analysis.
This circular region is also known as region of interest (ROI).
Further analysis procedure is the same as given in \citet{Prince et al. (2018)}.
I have analyzed the \textit{Fermi}-LAT data for 3C 279 from Nov 2017 to Jul 2018 and found that source has shown three significant
flares in these nine months.

\begin{figure*}
\vspace{-15pt}
 \centering
 \includegraphics[scale=0.45]{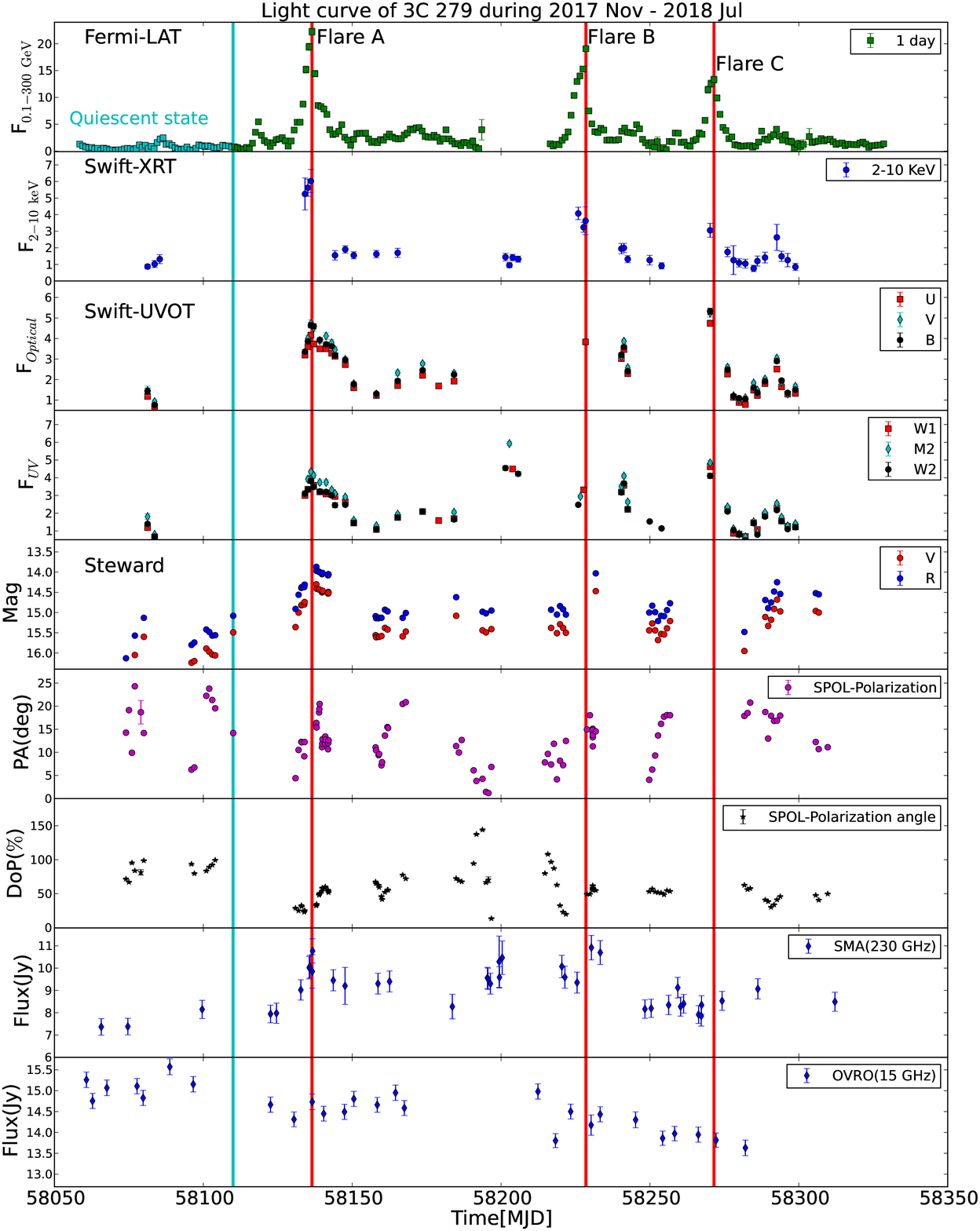}
 \vspace{-50pt}
 \caption{Multi-wavelength light curve of 3C 279 from November 2017 to July 2018. The $\gamma$-ray flux are presented in units of 10$^{-6}$ ph cm$^{-2}$ s$^{-1}$,
 and X-ray/UV/Optical fluxes are in units of 10$^{-11}$ erg cm$^{-2}$ s$^{-1}$. The vertical cyan color line separates the quiescent state and flaring state.
 The red vertical lines corresponding to each gamma-ray flares are drawn to recognised the flare in all wavebands.}
\end{figure*}

\subsection{Swift-XRT/UVOT}
Swift-XRT/UVOT is the space-based telescope which observes the galactic as well as extragalactic sources in X-rays, Opticals, and UV simultaneously.
Fermi detected blazars could also be observed by Swift-XRT/UVOT telescope as a monitoring program as well as a time of opportunity (ToO) program. 
The blazar 3C 279 was observed by Swift-XRT/UVOT when it was flaring during 2017-2018, and the details of the observations are present in Table 1.
I processed the XRT data by using the task `\textit{xrtpipeline}' version 0.13.2, which produces the cleaned event files for each observation.
While re-processing the raw data, I have used the latest calibration files (CALDB version 20160609).
The cleaned event files are produced only to the Photon Counting (PC) mode observations. 
The task `\textit{xrtpipeline}' is used to select the source and the background region.
Circular regions of radius 20 arcseconds around the source and slightly away from the source (fewer photon counts) are 
 chosen for the source and the background regions, respectively.
The X-ray spectra were extracted in '\textit{xselect}' and used as input spectra in '\textit{Xspec}' for modeling.
A simple power-law model with the galactic absorption column density $n_H$ = 1.77$\times$10$^{20}$ 
cm$^{-2}$ \citep{Kalberla et al. (2005)} is used to model the XRT spectra.

Simultaneous to XRT,
the Swift Ultraviolet/Optical Telescope (UVOT, \citealt{Roming et al. (2005)}) has also observed 3C 279 in all the available six filters U, V, B, W1, M2, and W2. 
The image is extracted by selecting a circular region of 5 arcsecond around the source and a circular region of 10 arcsecond
away from the source for the background. The task 'uvotimsum' is used to sum the multiple observations in the same filter at the
same epoch, and further, the task `uvotsource' is used to extract
the source magnitudes and fluxes. Magnitudes are corrected for galactic extinction using R$_{\rm V}$ = A$_{\rm V}$/E(B-V) = 3.1 and E(B-V) = 0.025
\citep{Schlafly and Finkbeiner (2011)} and converted into flux
using the zero points (\citealt{Breeveld et al. (2011)}) and conversion factors (\citealt{Larionov et al. (2016)}). 

\begin{table}
\centering
\caption{Table shows the log of the observations from Swift-XRT/UVOT telescope during the flaring state (MJD 58050 -- 58350).}
 \begin{tabular}{cc p{1cm}}
 \hline
 & \\
  Obs-ID & Exposure (ks)\\
 & \\
 \hline
 & \\
  00035019201 & 1.9\\
  00035019203 & 1.8\\
  00035019204 & 2.0\\
  00035019206 & 0.5\\
  00035019210 & 0.9\\
  00035019211 & 0.9\\
  00035019213 & 0.6\\
  00035019214 & 0.9\\
  00035019218 & 1.1\\
  00035019219 & 1.0\\
  00035019220 & 1.2\\
  00035019221 & 2.1\\
  00035019222 & 1.7\\
  00035019224 & 1.8\\
  00035019225 & 1.6\\
  00035019227 & 2.5\\
  00035019228 & 0.2\\
  00035019229 & 0.7\\
  00035019230 & 2.0\\
  00035019231 & 2.0\\
  00035019232 & 2.0\\
  00035019233 & 2.1\\
  00035019234 & 1.5\\
  00035019235 & 0.8\\
  00035019236 & 1.6\\
  00035019237 & 0.5\\
  00035019238 & 1.5\\
  00035019239 & 1.5\\
  00035019240 & 1.5\\
  00035019241 & 1.1\\
  00035019242 & 1.1\\
  00030867052 & 1.0\\
  00030867054 & 1.4\\
  00030867055 & 0.4\\
  00030867056 & 1.1\\
  00030867057 & 1.0\\
  00030867058 & 1.0\\
  00030867059 & 1.0\\
  00030867060 & 1.1\\
  00030867061 & 0.5\\
  00030867062 & 1.2\\
  00030867063 & 0.6\\
  00030867064 & 1.1\\
 &  \\
 \hline
 
 \end{tabular}
 \label{Table:T1}
\end{table}

\subsection{Steward Optical Observatory}
Steward Optical Observatory is a part of the \textit{Fermi} multiwavelength support program. It provides the optical data for the LAT-monitored 
blazars and also measures the linear optical polarization.
Archival data from the Steward Optical Observatory, Arizona \citep{Smith et al. (2009)}\footnote{http://james.as.arizona.edu/~psmith/Fermi/}
has been used in this particular study.

3C 279 is being continuously monitored with the SPOL CCD Imaging/Spectrometer of the Steward Observatory.
Optical V-band and R-band photometric and polarimetric (degree of polarization and position angle) data is collected for the flaring period of 
3C 279 from November 2017 to July 2018.  

\subsection{Radio data at 15 and 230 GHz}
Owens Valley Radio Observatory (OVRO; \citet{Richards et al. (2011)} is also a part of \textit{Fermi} monitoring program. It monitors the Fermi
blazars by a 40-meter single disc antenna at a frequency of 15 GHz. More than 1800 Fermi blazars are being continuously monitored by OVRO twice a week.
3C 279 is one of them, and I have collected the radio data at 15 GHz from November 2017 to July 2018.

Submillimeter Array provided the 230 GHz data (SMA) from observer center database \citep{Gurwell et al. (2007)}. The data is collected for the period
of 9 months from November 2017 to July 2018. 

\section{Results and Discussions}
I have analyzed the \textit{Fermi}-LAT, Swift-XRT/UVOT data from November 2017 to July 2018 (MJD 58060 -- MJD 58330) and the archival data from other 
telescopes like OVRO, SMA, and Steward observatory are collected for the same period.
The multiwavelength data from all the above telescopes are used to study the flux and polarization variability, and it is discussed in this 
particular section.
I have also presented a correlation study among different wavelength during the outburst period (MJD 58060 -- MJD 58330). 
A single zone emission model is chosen to perform the multiwavelength SED modeling.

\subsection{Multiwavelength Light Curves}
Blazar 3C 279 is known for its chaotic variability and active flaring behavior across the entire electromagnetic (EM) spectrum. 
The source was reported to be in a flaring state across the entire EM spectrum between 2017-2018.  
The multiwavelength light curve of 3C 279 during the flaring episode MJD 58060 -- MJD 58330 is shown in Figure 1. The first panel shows the one day
binning of \textit{Fermi}-LAT data in the energy range of 0.1 - 300 GeV. Our analysis shows the source was in flaring state during end of 2017 to 
mid 2018. The high activity started at the end of 2017 and followed by a flaring state defined as ``Flare A''.
After ``Flare A'', the source was observed for around two months in a low state with small fluctuations in the flux value. 
3C 279 again went to higher state and spent around two weeks in flaring state labeled as ``Flare B''. After one month period of ``Flare B'', the flux again
started rising, and source attained a full flaring state labeled as ``Flare C". The red bold vertical lines are drawn in Figure 1 corresponding to each
gamma-ray flare. Just before the ``Flare A'', the source has been observed in a long low flux state. 
I have chosen 50 days period between MJD 58060--58110 when the source flux is very low and constant over a long time. 
This period defined as ``quiescent state'' showed by cyan color data points in Figure 1 and separated 
by flaring state by a cyan color vertical line. The average flux observed during this period is 0.73$\times$10$^{-6}$ ph cm$^{-2}$ s$^{-1}$.

Swift-XRT/UVOT has also monitored the source during the gamma-ray outburst. The X-rays, optical, and UV light curves are shown in panels
2, 3, $\&$ 4 of Figure 1. 3C 279 has gone through the strong X-ray flaring episode corresponding to each gamma-ray flare. In Optical and UV 
band 3C 279 has shown the flaring behavior corresponding to ``Flare A'' and ``Flare C'', while ``Flare B'' of gamma-ray is missing in Optical and 
UV because of the unavailability of UVOT observations.

The 3C 279 has also been monitored by Steward Observatory in optical V and R band. In panel 5 of Figure 1, I have plotted the Steward data, and it can
be seen that the source is highly variable in both the bands. The ``Flare A'' of gamma-ray is followed by the flare in both V and R band. 
The ``Flare B'' is observed close in time at optical (Steward Observatory) and gamma-ray, while ``Flare C'' is not seen in Steward Observatory 
because of poorly sparsed V and R band data points.

In the 6th and 7th panel of Figure 1, I have plotted the optical degree of polarization (DoP) and polarization angle (PA) from Steward Observatory.
Huge variations are seen during the flaring period. DoP varies from 4$\%$--22$\%$ in 12 days of span (MJD 58130--58142) and on the other hand a slow
change is seen in polarization angle from 25$\degree$--60$\degree$.

The radio light curve from SMA and OVRO observatory at 230 and 15 GHz are shown in the last two panel of Figure 1. The radio data at 230 GHz 
from SMA observatory shows that the source is variable in this energy band, and high state radio flux has been noticed 
during ``Flare A'' and ``Flare B''.
It is observed that there are not much variations in OVRO flux during the flaring period of gamma-ray/X-ray/Optical/UV. 

\subsection{Variations in Gamma-ray}
In Figure 2, the gamma-ray flares are plotted separately along with the corresponding photon spectral index. The ``Flare A'' is shown in the
leftmost panel of Figure 2. 
The source starts showing the activity at MJD 58115 with a small rise in flux. The small fluctuations continued until the source went to the
higher state, where the flux rose above the quiescent state flux value within the 10 days of time interval (MJD 58129--58140). The flux at 
MJD 58129.5 is 1.89$\times$10$^{-6}$ ph cm$^{-2}$ s$^{-1}$ and almost after 7 days the flux rose up to 22.24$\times$10$^{-6}$ ph cm$^{-2}$ s$^{-1}$
at MJD 58136.5 and the photon spectral index became harder changing from 2.43 to 2.19, respectively. Just after three days from the peak, the flux
started decreasing and within 10 days, it attained the low flux state at MJD 58149.5 with a flux of 1.08$\times$10$^{-6}$ ph cm$^{-2}$ s$^{-1}$ and
spectral index 2.31.

The middle panel of Figure 2 represents the light curve of ``Flare B''. The activity was found to commence at MJD 58215 just after two months of 
``Flare A''. The flux started rising very slowly from MJD 58216.5 with flux 1.35$\times$10$^{-6}$ ph cm$^{-2}$ s$^{-1}$ and in span of two weeks
it achieved the maximum flux value of 19.06$\times$10$^{-6}$ ph cm$^{-2}$ s$^{-1}$ at MJD 58228.5.
The spectral index became harder from 2.84 to 2.04.
Within one day the flux dropped from (19.06--7.49)$\times$10$^{-6}$ ph cm$^{-2}$ s$^{-1}$ and then the source went through small fluctuations in flux
for 20 days before reaching the quiescent state. The quiescent state flux noticed at MJD 58249.5 is 
1.00$\times$10$^{-6}$ ph cm$^{-2}$ s$^{-1}$.

The ``Flare C'' is shown in the rightmost panel of Figure 2. As soon as ``Flare B'' ends, the flux started fluctuating again above the quiescent state
flux value (0.27$\times$10$^{-6}$ ph cm$^{-2}$ s$^{-1}$) and it continues till one week and shows a little rise in flux (3.36$\times$10$^{-6}$ ph cm$^{-2}$ s$^{-1}$).
After spending two days in a high flux state, the source comes back to low flux state, and after five days it again rises to higher flux state. The maximum
flux achieved during high flux state is 13.31$\times$10$^{-6}$ ph cm$^{-2}$ s$^{-1}$ at MJD 58271.5.

One day bin light curve shown in Figure 2 are used to estimate the variability time, which can be defined by the following equation
\citep{Zhang et al. (1999)}, 
\begin{equation}
\centering
t_{var} =  \frac{F_1 + F_2}{2} \frac{t_2 - t_1}{|F_2 - F_1|}
\end{equation}
where \textit{$F_1$} and \textit{$F_2$} are the fluxes measured at time $t_1$ and $t_2$. The above equation (1) is used to scan the 1 day bin light
curve (Figure 2) to find the variability time. 
The shortest variability time is found to be order of days in all the three flares and the details are shown in Table 2.
In Figure 3, the gamma-ray flux and the photon spectral index during the flaring period are plotted together along the x and y-axis, respectively. A 
``harder-when-brighter'' trend of source is seen here. Similar trend is also seen previously for 3C 279 by
 \citet{Hayashida et al. (2012)}, \citet{Hayashida et al. (2015)}, $\&$ \citet{Paliya (2015a)}. The spectral hardening during the flaring state
 can predict the possibility of detection of high energy photons and consequently can shift the IC peak of the SED to higher energy. A strong
 correlation between spectral hardening during flare and detection of high energy photons are shown by many authors (\citealt{Britto et al. (2016)};
 \citealt{Shah et al. (2019)}).

%\textbf{The average photon spectral index is estimated as 2.26$\pm$0.01, which is softer than the 
%2.19$\pm$0.03, the observed index value corresponds to the highest flux (22.24$\pm$0.70)$\times$10$^{-6}$ ph cm$^{-2}$ s$^{-1}$.}
\begin{table}
\centering
\caption{Table shows the variability time estimated from equation (1) for all the different flares. The flux F$_1$ and F$_2$ are in units of 
 10$^{-6}$ ph cm$^{-2}$ s$^{-1}$ and t$_1$  $\&$ t$_2$ are in MJDs .}
 \begin{tabular}{cccccc p{1cm}}
 \hline \\
 Flares & F$_1$ & F$_2$ & t$_1$  & t$_2$  & t$_{var}$ (days) \\
 \hline \\
% Fermi-LAT &&&& & (hr) \\
 Flare A & 1.89  & 3.91 & 58129.5 & 58130.5 & 1.43$\pm$0.16    \\
 Flare B & 19.06  & 7.49 & 58228.5 & 58229.5 & 1.14$\pm$0.03    \\
 Flare C & 5.73  & 11.44 & 58268.5 & 58269.5 & 1.50$\pm$0.11    \\
 \hline \\
\end{tabular}
 \end{table}
 
\begin{figure*}
\centering
 \includegraphics[scale=0.375]{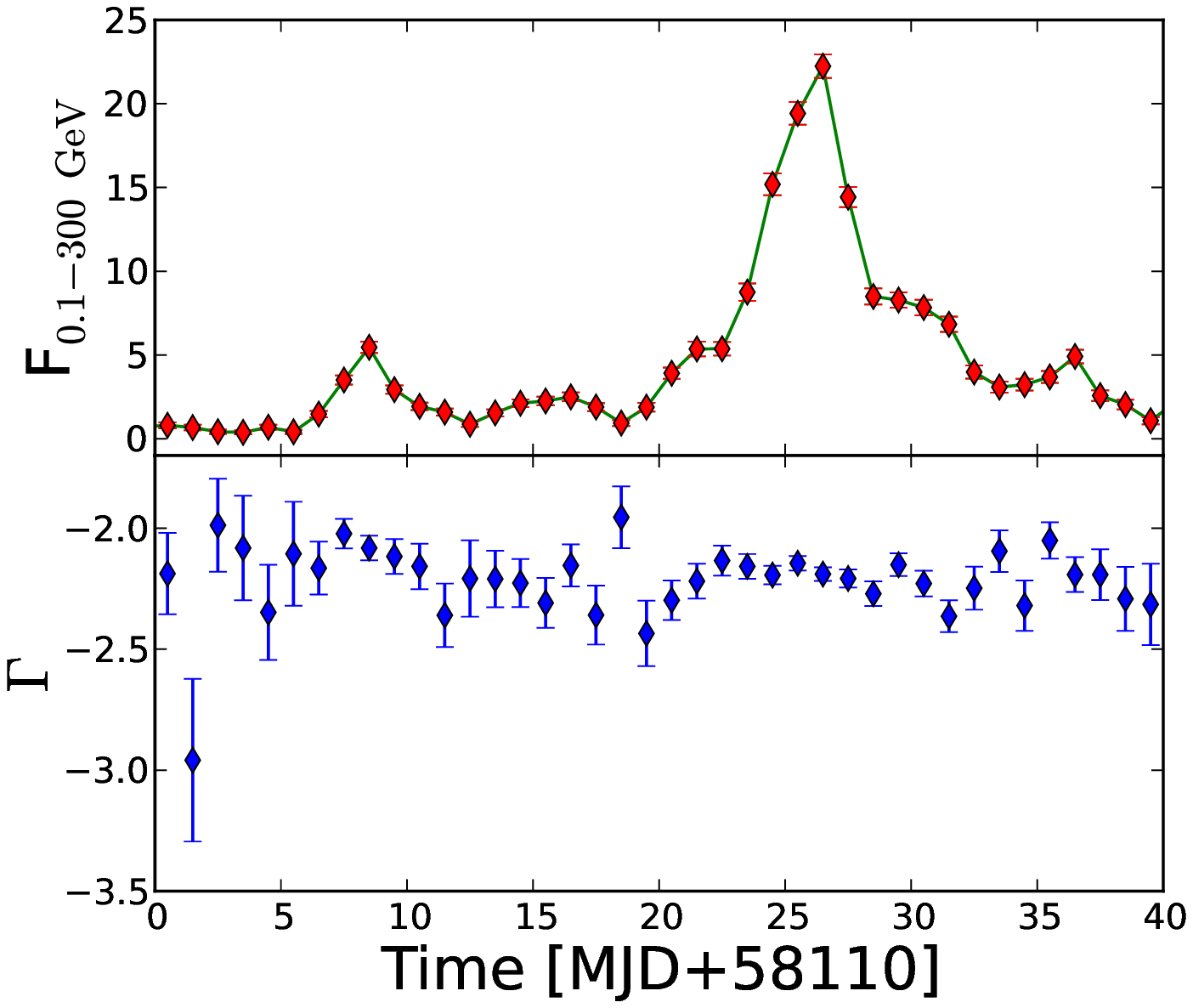}
 \includegraphics[scale=0.375]{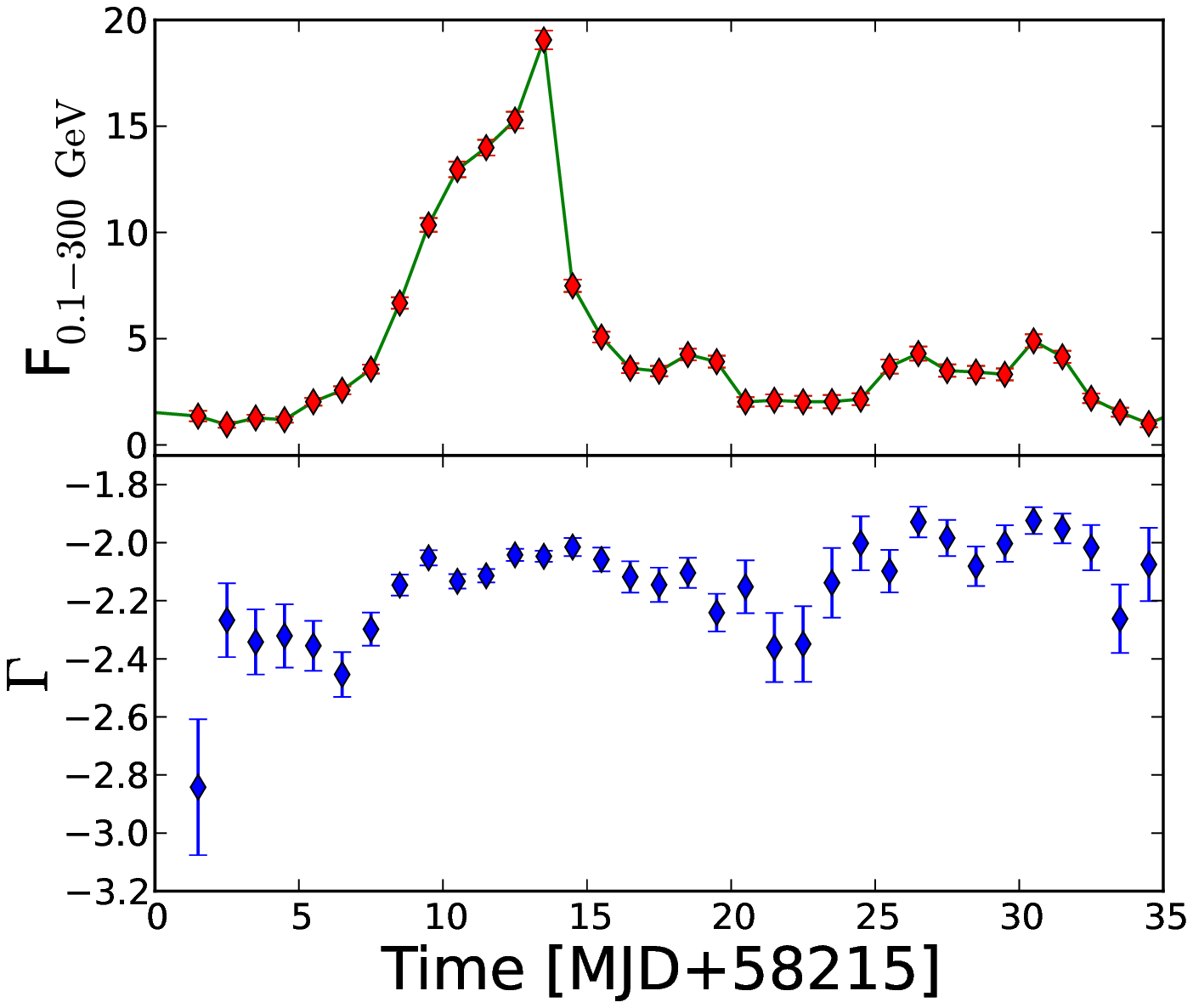}
 \includegraphics[scale=0.375]{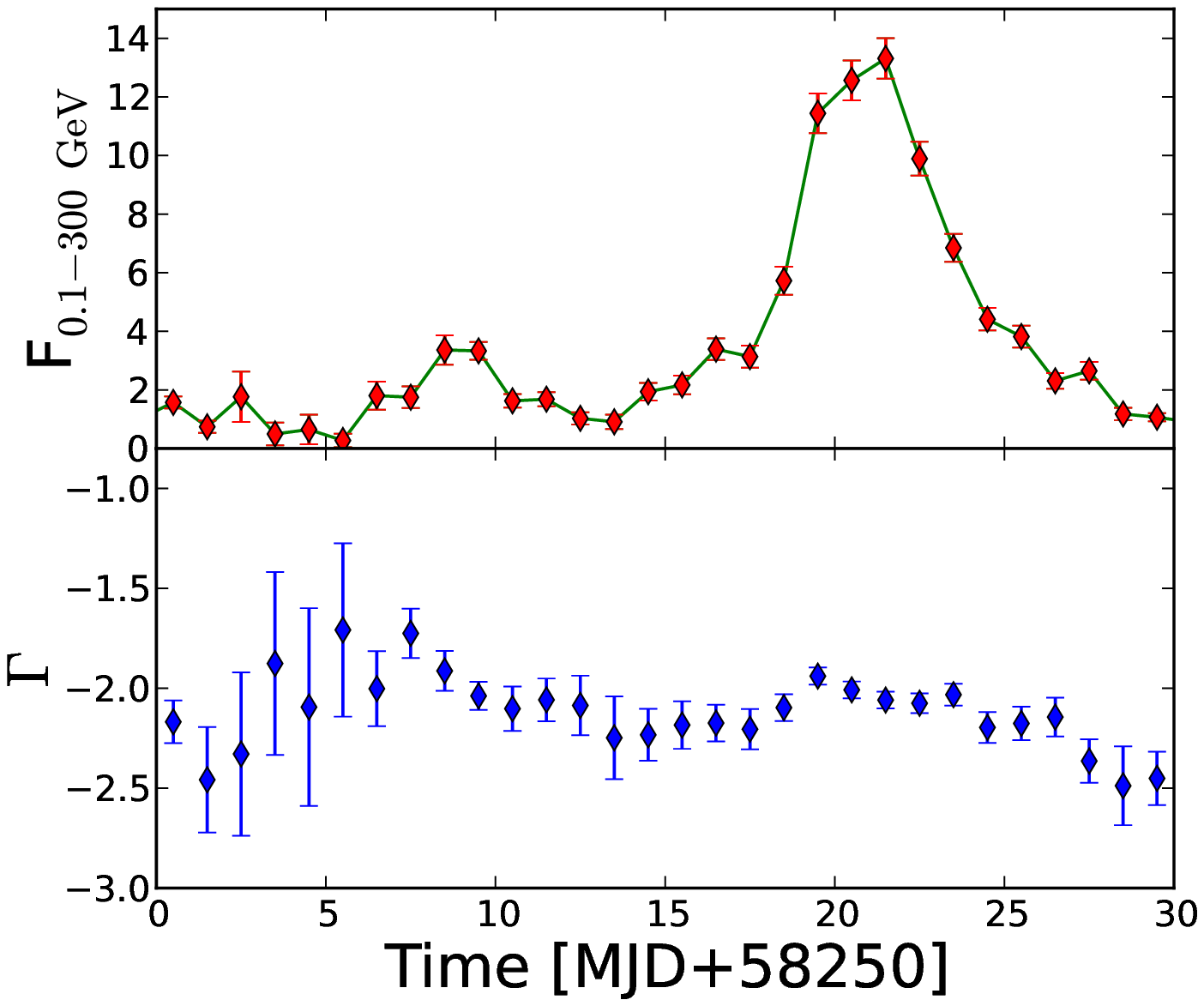}
 \caption{Gamma-ray light curve for all the three flares separately with corresponding photon spectral index. The y-axis of the upper panels are
 in units of 10$^{-6}$ ph cm$^{-2}$ s$^{-1}$.}
 \end{figure*}
 
\begin{figure}
 \centering
 \includegraphics[scale=0.35]{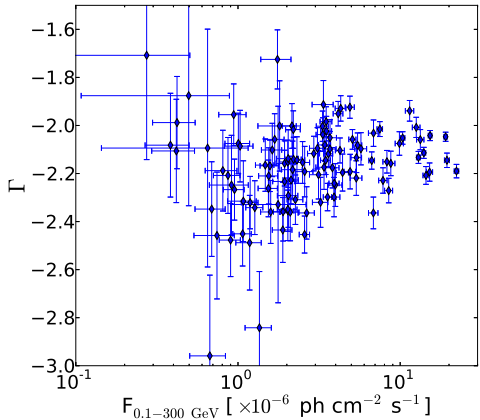}
 \caption{The gamma-ray photon spectral index corresponding to flux observed during flare shows a ``harder-when-brighter'' trend.} 
 %The x-axis is in units of ph cm$^{-2}$ s$^{-1}$.}
\end{figure}

\subsection{Variations in X-ray}
\begin{figure}
\centering
 \includegraphics[scale=0.45]{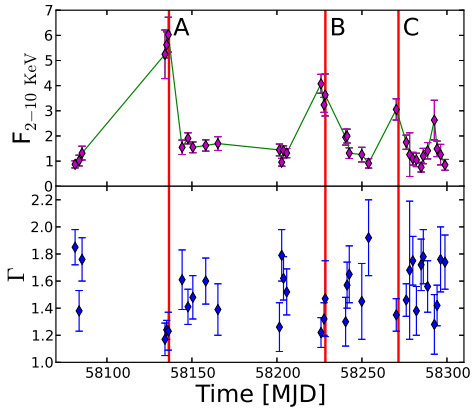}
 \caption{X-Ray light curve for all the observed flares. The fluxes are in units of 1.0$\times$10$^{-11}$ erg cm$^{-2}$ s$^{-1}$. Lower panel
represent the corresponding photon spectral index. A ``harder-when-brighter'' trend is also seen here.}
\end{figure}

Swift-XRT light curve of all the flares for 2-10 keV is shown in Figure 4 along with the photon spectral index. The observations done by Swift-XRT
are poorly sparsed, and as a result, there is no clear indication of rising and decaying part of the flares. The maximum flux achieved during ``Flare A"
in X-ray is (6.03$\pm$0.70)$\times$10$^{-11}$ erg cm$^{-2}$ s$^{-1}$ with spectral index 1.23$\pm$0.14. 
After around two months period of ``Flare A", ``Flare B" is 
observed in X-ray at the
same time as the gamma-ray flare. The maximum flux achieved in X-ray during this period is (4.07$\pm$0.37)$\times$10$^{-11}$ erg cm$^{-2}$ s$^{-1}$ 
and the spectral index
noticed as 1.22$\pm$0.11. The last flare in X-ray, i.e. ``Flare C", is followed by a small fluctuations at the end. The highest
flux observed during ``Flare C" is (3.06$\pm$0.42)$\times$10$^{-11}$ erg cm$^{-2}$ s$^{-1}$, and the corresponding spectral index found to be 1.35$\pm$0.12. 

Figure 4 reveals ``harder-when-brighter'' behavior in X-ray as well, which is also reported by \citet{Hayashida et al. (2012)} $\&$ \citet{Hayashida et al. (2015)}. 
The average photon spectral index is estimated as 1.52$\pm$0.03, that is softer 
than the spectral index observed at highest flux i.e. 1.23$\pm$0.14 during the ``Flare A". Most of the contribution of soft X-ray goes to 
synchrotron peak and the lower part of the SSC/IC peak. The ``harder-when-brighter'' trend in X-ray can be interpreted as the increase in the
SSC emission and can also shift the SSC peak towards the higher energy. Increase in the SSC/IC emission could be probably due to an increase of the
accretion rate.

The variability time is a tool to find out the size and location of
the emission region. In the case of X-ray emission, the light curve is poorly sparsed and so can not be used to estimate the variability time in X-ray.
The location of the X-ray emission region can be estimated indirectly by the correlations study between gamma-ray and X-ray emissions. The details
of the correlations study are shown in section 3.6.

\subsection{Spectral Ananlysis}
The $\gamma$-ray spectral energy distributions (SEDs) have been generated for all the three flares and one quiescent state between 0.1 -- 300 GeV,
from the likelihood analysis.
The spectral data points are plotted in Figure 5, I have used three different spectral models to fit the SEDs data points. The spectral models are 
Power-law (PL), Logparabola (LP), and simple broken Power-law (BPL), and corresponding expressions are given in \citet{Prince et al. (2018)}. 
The presence or absence of curvature in the gamma-ray SED plays a crucial role in constraining the location of the emission region.
A break in the gamma-ray spectrum is expected when the emission region is within the broad line region (BLR), because of photon-
photon pair production. Earlier study by \citet{Liu and Bai (2006)} has shown that the BLR region is opaque to photons of energy $>$ 20 GeV
, and as a result, a curvature or break can be seen in the gamma-ray spectrum above 20 GeV. 
The curvature in the gamma-ray spectrum can be justified by estimating the TS$_{curve}$. According to \citet{Nolan et al. (2012)}
the TS$_{curve}$ can be defined as TS$_{curve}$ = 2(log L(LP/BPL) - log L(PL)), where L represents the likelihood function.
The model parameters and the value of TS$_{curve}$ are mentioned in Table 3.
The model which have a large positive value of TS$_{curve}$ is considered as the best fit to the SEDs data points and it suggests the
presence of a spectral cut-off. From Table 3, it is very much clear that the log-parabola spectral model better explains the gamma-ray SED data. 
I have also noticed that the break energy found in broken power-law fit is constant irrespective of the different flares. This finding is also
consistent with the previous study on this source, by \citet{Paliya (2015a)}, and on 3C 454.3 by \citet{Abdo et al. (2011)}. 
There is also an alternative way to explain this curvature/cut-off. A cut-off in the energy spectrum can also 
occur when there is already a cut-off in the energy distributions of the particles.

A strong break is seen in the gamma-ray spectrum while fitting with BPL; similar break has also been seen before for other FSRQ like 3C 454.3 by 
\citet{Abdo et al. (2011)}
and also for 3C 279 during flare of 2014 and 2015 (\citealt{Paliya (2015a)}; \citealt{Paliya et al. (2015b)}). In table 3, during all the flares 
the BPL photon index ($\Gamma_1$) before the break (E$_{break}$ = 1 GeV) is $\leq$ 2, which suggest an increasing slope, and after the break the 
photon spectral index ($\Gamma_2$) is $\geq$ 2, indicates a falling spectrum. We know that the
gamma-ray energy spectrum is governed by inverse Compton (IC) scattering and my result shows it peaks around 1 GeV which lies in the
energy range of LAT (0.1-300 GeV). Since the break energy is almost constant for all the flares, it is possible that the observed shape is a
reminiscence of the electron energy distribution of the emitting electrons.

\begin{figure*}
\begin{center}
 \includegraphics[scale=0.35]{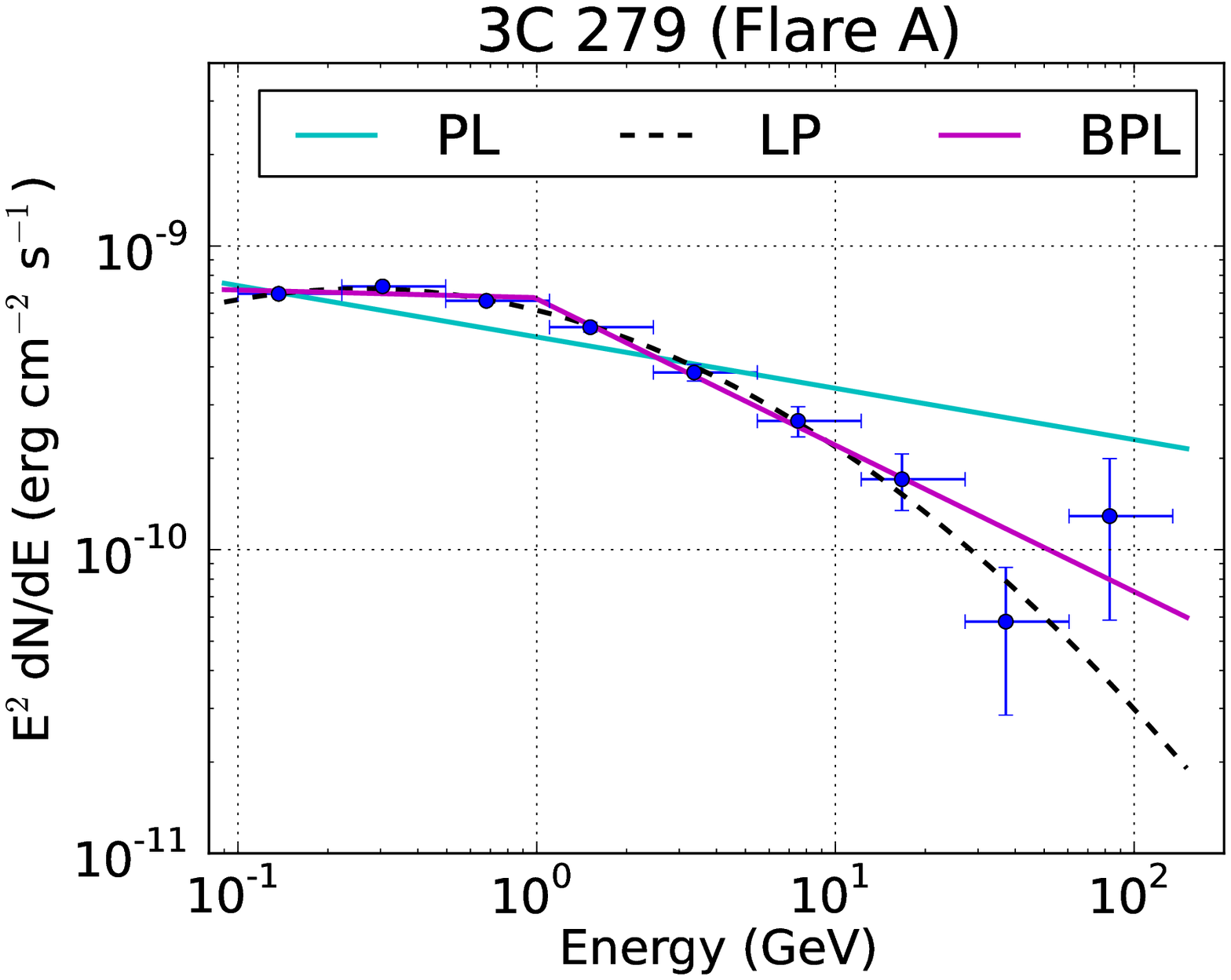}
 \includegraphics[scale=0.35]{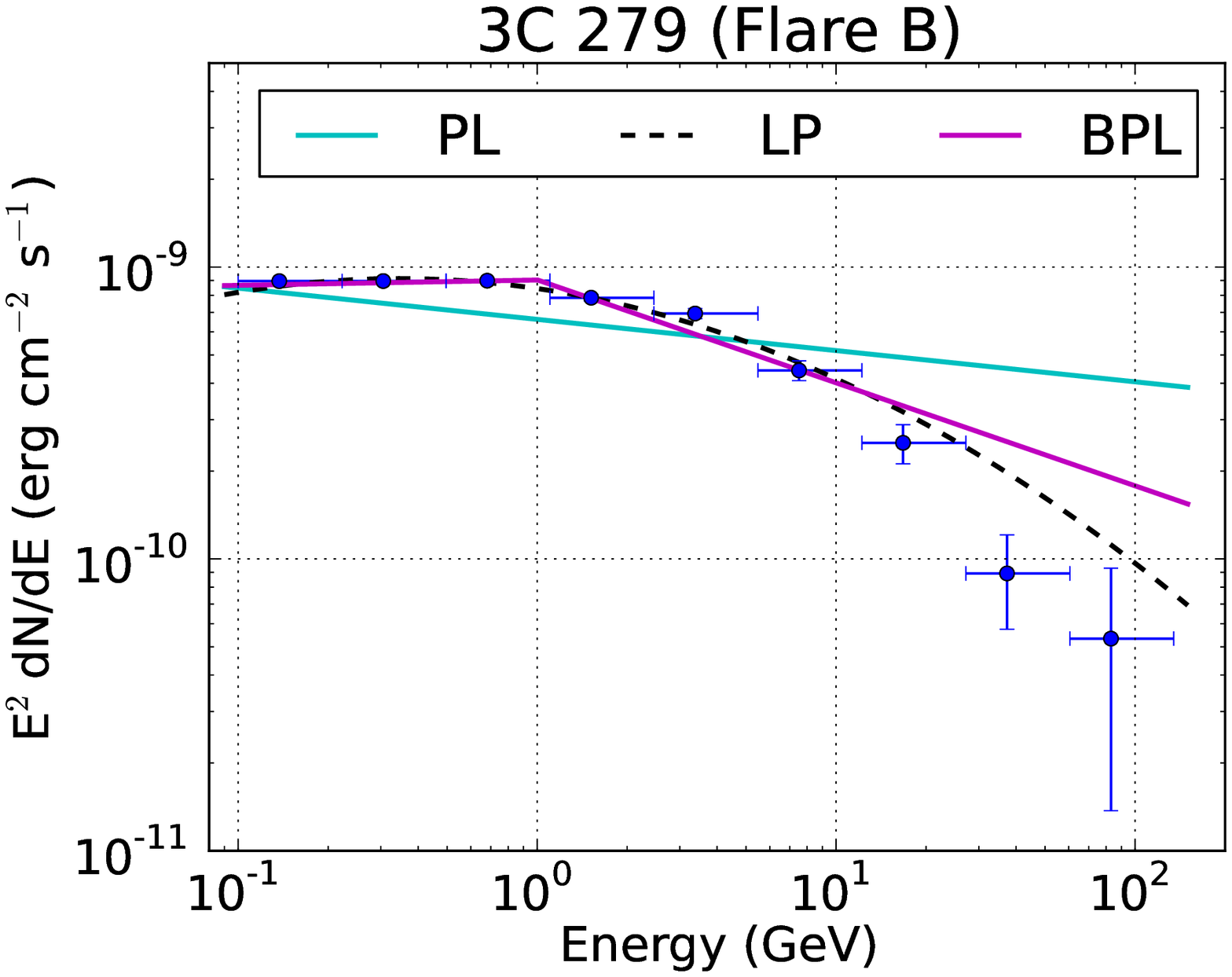}
 \includegraphics[scale=0.35]{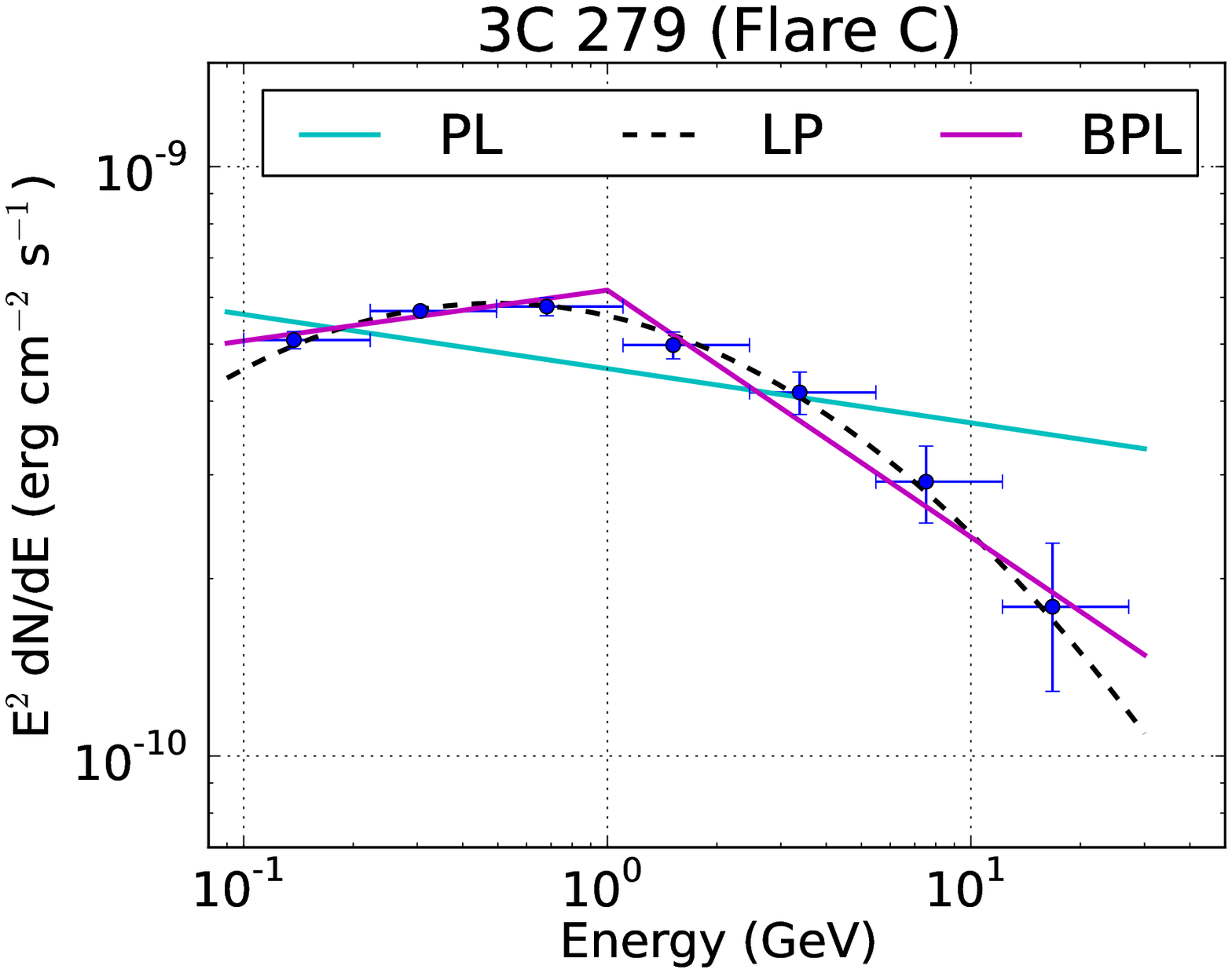}
 %\begin{center}
 \includegraphics[scale=0.35]{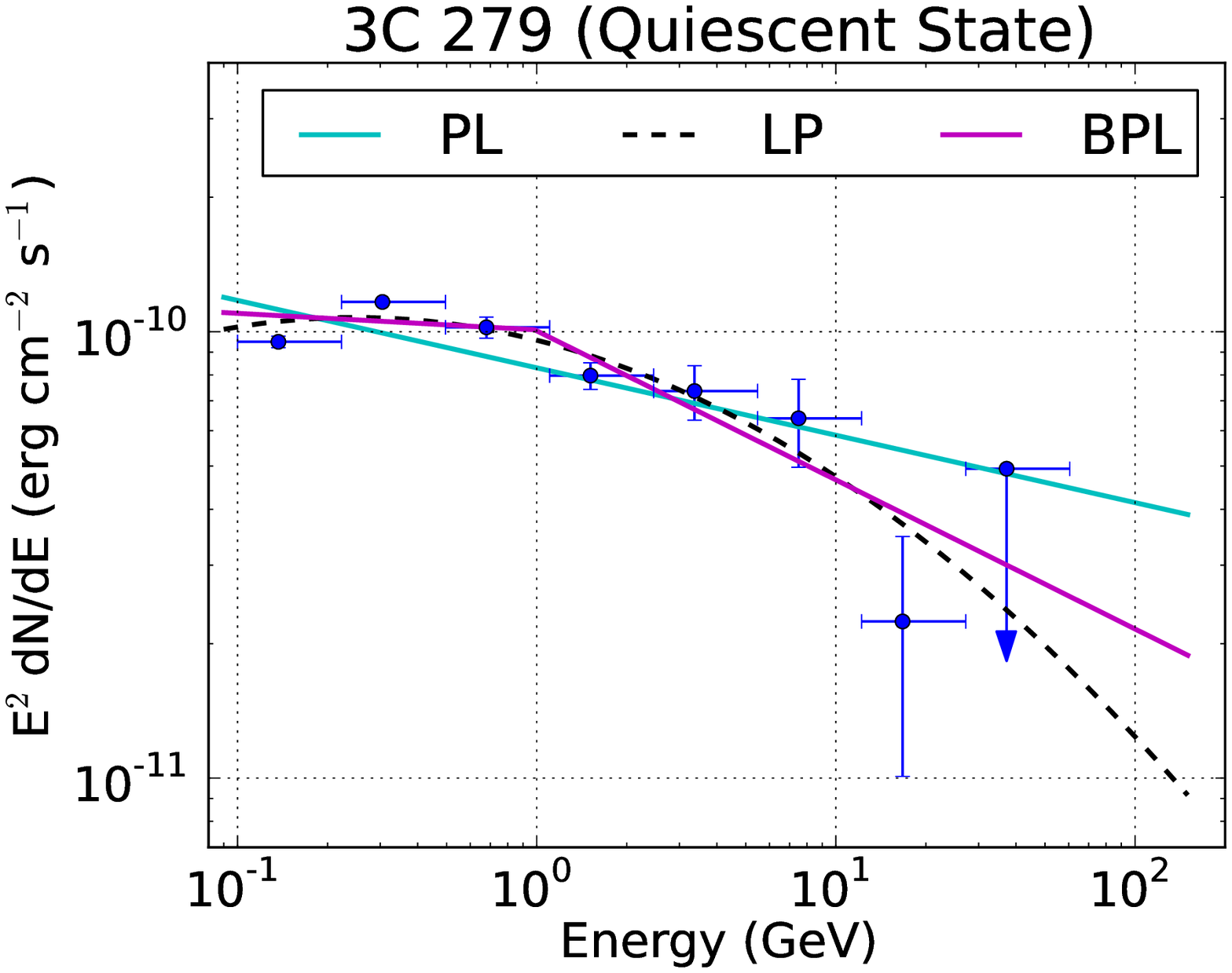}
\end{center}
 \caption{The gamma-ray SEDs produced for all the flares and quiescent state. The data points are fitted with three different models and the modeled
 parameters are shown in Table 3.}
\end{figure*}

\begin{table*}
\begin{center}
 \caption{Parameters obtained from the spectral analysis fit, for the different models PL, LP, and BPL, by using 
 the Likelihood analysis method. TS$_{curve}$ is estimated with respect to the TS value of the PL fit.}
 \begin{tabular}{c c c c c c c}
 \toprule
  &&PowerLaw (PL)&&&&  \\
  \midrule
Activity  &  F$_{0.1-300}$ $_{\rm{GeV}} $ & $\Gamma$ & && TS & TS$_{curve}$ \\[0.5ex]
& ($10^{-6}$ ph cm$^{-2}$ s$^{-1}$)&&&&  \\
\midrule
Quiescent state&0.67$\pm$0.02& 2.15$\pm$0.03 &- &- & 3625.24 \\ 
Flare A& 4.42$\pm$0.05 & 2.17$\pm$0.92  &- &-& 35533.08& - \\
Flare B& 5.59$\pm$0.05 &2.11$\pm$0.01 & - & - & 67293.68 & -   \\
Flare C& 3.47$\pm$0.06 &2.09$\pm$0.02 &- & - & 15163.31 & -  \\
\midrule
&&LogParabola (LP)&&&  \\
  &   & $\alpha$ & $\beta$& &   &  \\[0.5ex]
\midrule
Quiescent state&0.64$\pm$0.03 & 2.04$\pm$0.04 & 0.06$\pm$0.02 & - & 3653.27 & 28.03\\
Flare A& 4.24$\pm$0.05 & 2.05$\pm$0.02 &0.090$\pm$0.008 &-& 35759.77& 226.69  \\
Flare B& 5.39$\pm$0.05 & 1.99$\pm$0.01 &0.072$\pm$0.006 &-& 67311.92 & 18.24 \\
Flare C& 3.26$\pm$0.07 & 1.93$\pm$0.02 & 0.10$\pm$0.01 & -& 15282.46 &  119.15 \\
\midrule
&&Broken PowerLaw (BPL)&&&  \\
&   &$\Gamma_{1}$ & $\Gamma_{2}$ & E$_{break}$ & &   \\
\midrule
Quiescent state& 0.65$\pm$0.03 & 2.04$\pm$0.05 & 2.33$\pm$0.08 & 0.98$\pm$0.17 & 3633.04 & 7.8 \\
Flare A& 4.27$\pm$0.05 & 2.02$\pm$0.02  &2.48$\pm$0.03 &0.98$\pm$0.05 & 35750.45& 217.37 \\
Flare B& 5.42$\pm$0.05 & 1.98$\pm$0.01  &2.35$\pm$0.02 &1.00$\pm$0.03 & 67309.47& 15.79  \\ 
Flare C& 3.30$\pm$0.07 & 1.91$\pm$0.03  &2.42$\pm$0.05 &1.00$\pm$0.06 & 15263.59 & 100.28  \\
\bottomrule
\end{tabular}
\end{center}
\end{table*}

\subsection{Fractional variability (F$_{var}$)}
Blazars are a particular class of AGN that show strong chaotic variability at all frequencies. The variability is more evident during the flaring
period and the flare profiles mostly depend on the particle injection, particle acceleration, and energy dissipation in the jets of blazars.
The variability amplitude can be estimated by measuring all kind of jet parameters like the magnetic field in the jets, viewing angle of the jet,
particle density inside the jet, and finally the efficiency of the acceleration process involved within the jet.
The challenge to determine the variability amplitude across the energy band requires a good quality of data. 3C 279 is a well-observed source across
the entire electromagnetic spectrum, and that makes it possible to estimate the variability amplitude. The fractional root mean square (rms) variability
parameter (F$_{var}$) is a tool to determine the variability amplitude and that is introduced by \citet{Edelson and Malkan (1987)}; \citet{Edelson et al. (1990)}.
The relation given in \citet{Vaughan et al.(2003)} is used to determine the fractional variability (F$_{var}$),
\begin{equation} \label{8}
 F_{var} = \sqrt{\frac{S^2 - \sigma^2}{r^2}} \\
\end{equation}

\begin{equation}
 err(F_{var}) = \sqrt{  \Big(\sqrt{\frac{1}{2N}}. \frac{\sigma^2}{r^2F_{var}} \Big)^2 + \Big( \sqrt{\frac{\sigma^2}{N}}. \frac{1}{r} \Big)^2     } \\
\end{equation}

where, $\sigma^2_{XS}$ = S$^{2}$ -- $\sigma^2$, is called excess variance, S$^{2}$ is the sample variance, $\sigma^2$ is the mean square uncertainties
of each observation and $r$ is the sample mean. 

\begin{table}
\centering
\caption{Fractional variability in various wavebands are estimated for the time interval MJD 57980 to 58120.}
\begin{tabular}{c c c c c}
 \hline
 \\
 Waveband  & F$_{var}$ & err(F$_{var}$) \\
 \\
 \hline
 
 $\gamma$-ray &  1.201 & 0.008 \\ 
 
 X-ray & 0.660 & 0.035 \\

 U & 0.524& 0.008 \\
 
 B & 0.520& 0.006 \\

 V & 0.509& 0.008 \\
 
 W1 & 0.548& 0.008 \\
 
 M2 & 0.580& 0.007 \\
 
 W2 & 0.557& 0.007 \\
 
 OVRO (15 GHz) & 0.032 & 0.002 \\
 
 SMA (230 GHz) & 0.089 & 0.010 \\
 
 \hline
\end{tabular}
\label{Table:TA}
\end{table}

The fractional variability estimated across the entire electromagnetic spectrum is shown in Figure 6, and the values are presented in Table 4.
It is seen that the source variability is more than 100$\%$ in $\gamma$-ray, followed by X-ray at more than 60$\%$, and then followed by UV and 
Optical, where the source variability is more than 50$\%$. In radio at 15 and 230 GHz, the source is very less variable (less than 10$\%$) during
this particular gamma-ray flaring period.
The F$_{var}$ is 1.201 in $\gamma$-ray, 0.580 in UVM2-band, 0.520 in optical U-band and 0.032 in radio at 15 GHz, and 0.089 at 230 GHz. 
It is found that F$_{var}$ is increasing with energy, suggesting that a large number of particles are injected in the jet, resulting in high
energy emission. Similar behavior of fractional variability is also seen for other FSRQ like Ton 599 by \citet{Prince (2019)},
where he found a trend of large fractional variability towards higher energies. This is not always the case, and an opposite trend was also
noticed in the past. An opposite trend is reported by \citet{Bonning et al. (2009)}, where
variability amplitudes decrease towards shorter wavelength (IR, Optical, and UV),
which suggests the presence of steady thermal emission from the accretion disk.

\begin{figure}
\centering
 \includegraphics[scale=0.355]{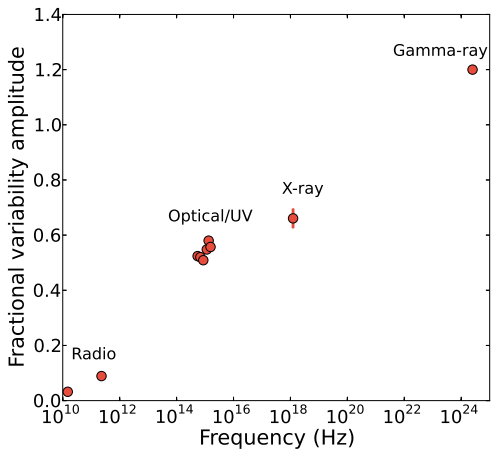}
 \caption{Fractional variability for various wavebands are plotted with respect to their frequency.}
\end{figure}

\subsection{Correlation Studies}
In Figure 1, all the light curves are plotted together, and the visual inspection suggests
that the flares in $\gamma$-ray, X-ray, Optical, and UV band are mostly correlated. The radio light curve at 15 GHz does not
show any flare for this particular period, while the light curve at 230 GHz shows a rise in flux corresponding to ``Flare A" and 
``Flare B". The gamma-ray ``Flare C" detected in X-rays, and Optical/UV have not seen in radio at 230 GHz. 
The cross-correlation study of flux variations in different energy bands are presented in this section. The correlations study help to reveal
the location of the emission region in different energy bands with respect to each other. A strong and good correlation between two energy emissions
indicates their co-spatial origin. A positive and negative time lag between two energy bands with strong correlation coefficient
suggest the different emission region for different waveband emissions and the time lag can be used to estimate the separation between them
\citep{Fuhrmann et al. (2014)}.
The correlation study is done by using the \textit{ discrete correlations function (DCF)} formulated by 
\citet{Edelson and Krolik (1988)}. 
When the two light curves, suppose LC1 and LC2 are correlated then a positive time lag between these two light curves implies that the LC1 is
leading the LC2, and a negative time lag means the opposite.
The results of \textit{DCFs} are shown in Figure 7, Figure 8 $\&$ Figure 9 for ``Flare A" $\&$ ``Flare C", for various waveband
combinations.

A cross-correlation study between two light curves
(different energy band) by \textit{DCF} needs sufficient data points in each light curve. The ``Flare B" fails to fulfil this criteria since
it has good quality data points only in the gamma-ray light curve. It is therefore not possible to cross-correlate two different energy band for ``Flare B".
Significant correlation has been noticed between gamma-ray and optical band (U, B, \& V) emissions with zero time lag (within the time bin 2.5 days).  
The correlation coefficient for ``Flare A'' is found to be 0.50$\pm$0.15, 0.53$\pm$0.17, 0.83$\pm$0.18 and for ``Flare C'' it is 0.92$\pm$0.44, 
0.93$\pm$0.44, 0.91$\pm$0.44 in U, B, \& V band, respectively.
The cross-correlation between gamma-ray and UV band (W1, M2, \& W2) emission are very similar to the gamma-ray vs optical band 
emission. The correlation coefficients for ``Flare A'' and ``Flare C'' are 0.45$\pm$0.15, 0.83$\pm$0.17, 0.82$\pm$0.17 and 0.92$\pm$0.44, 
0.90$\pm$0.43, 0.89$\pm$0.43 in W1, M2, \& W2 band, respectively.
A strong correlation with zero time lag is observed between gamma-ray and X-ray emission for 3C 279. In author's knowledge this is the first time 
the source has shown strong correlation with zero time lag between gamma-ray and X-ray.
The value of correlation coefficients for ``Flare A'' and ``Flare C'' are noted as 0.87$\pm$0.16 and 0.97$\pm$0.36 respectively, at zero time lag (within
the time bin). 
The significance of the DCF peaks are also estimated by simulating the 1000 gamma-ray light curves. To simulate the gamma-ray light curves, I have
used method mentioned in \citet{Emmanoulopoulos et al. (2013)}. Further, I cross-correlate the simulated gamma-ray light curves with the observed
light curves in different wavebands and finally a DCF distribution with time lag is calculated. At each time lag 95$\%$ significance was calculated,
which are shown in cyan color in Figure 7, Figure 8, and Figure 9.

%%%%%%%%%%%%%%%%%%%%%%%%%%%%%%%%%%%%%%%%%%%%%%%%%%%%%%%%%%%%%%
\begin{figure*}
\centering
 \includegraphics[scale=0.21]{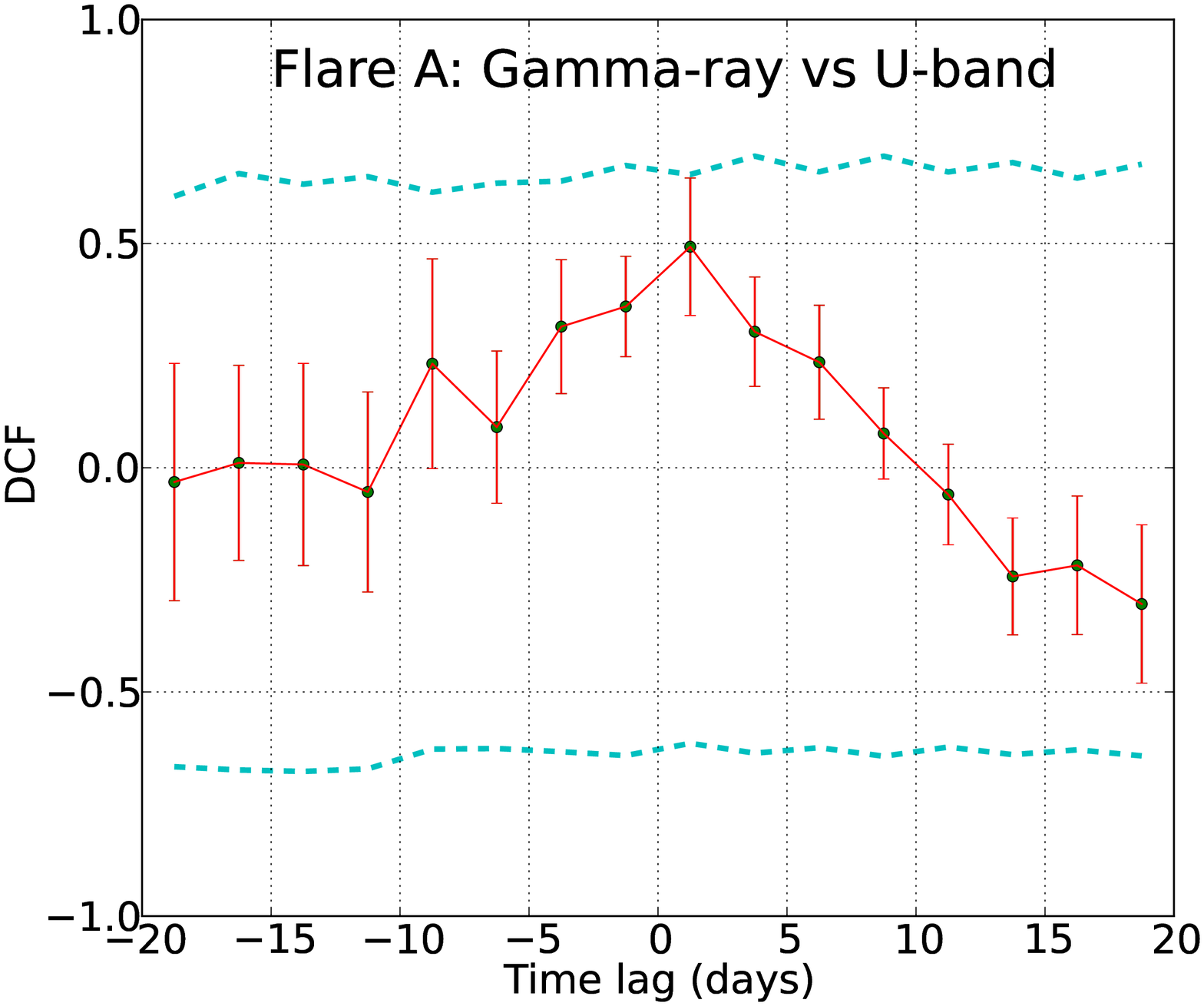} 
 \includegraphics[scale=0.21]{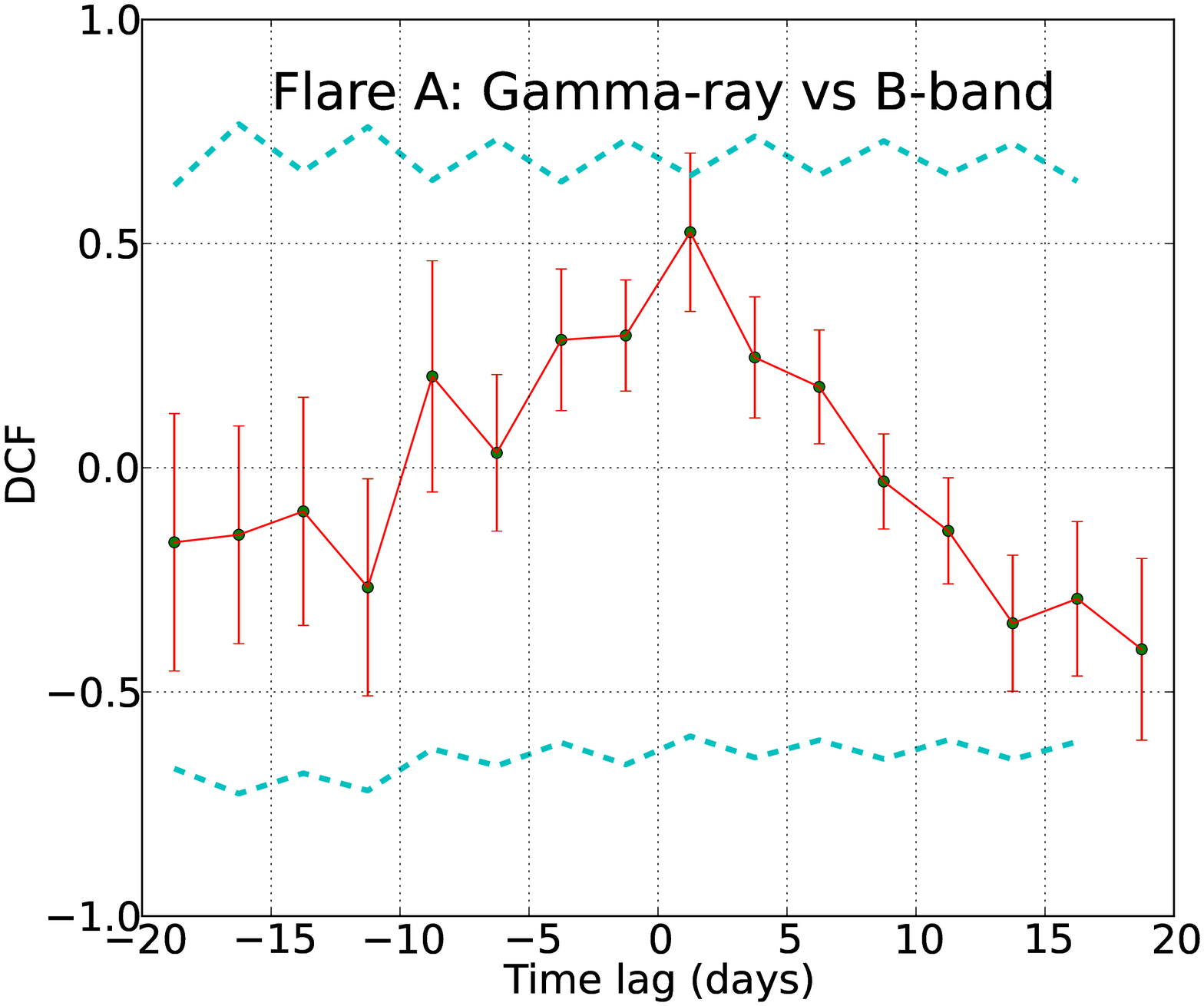}
 \includegraphics[scale=0.21]{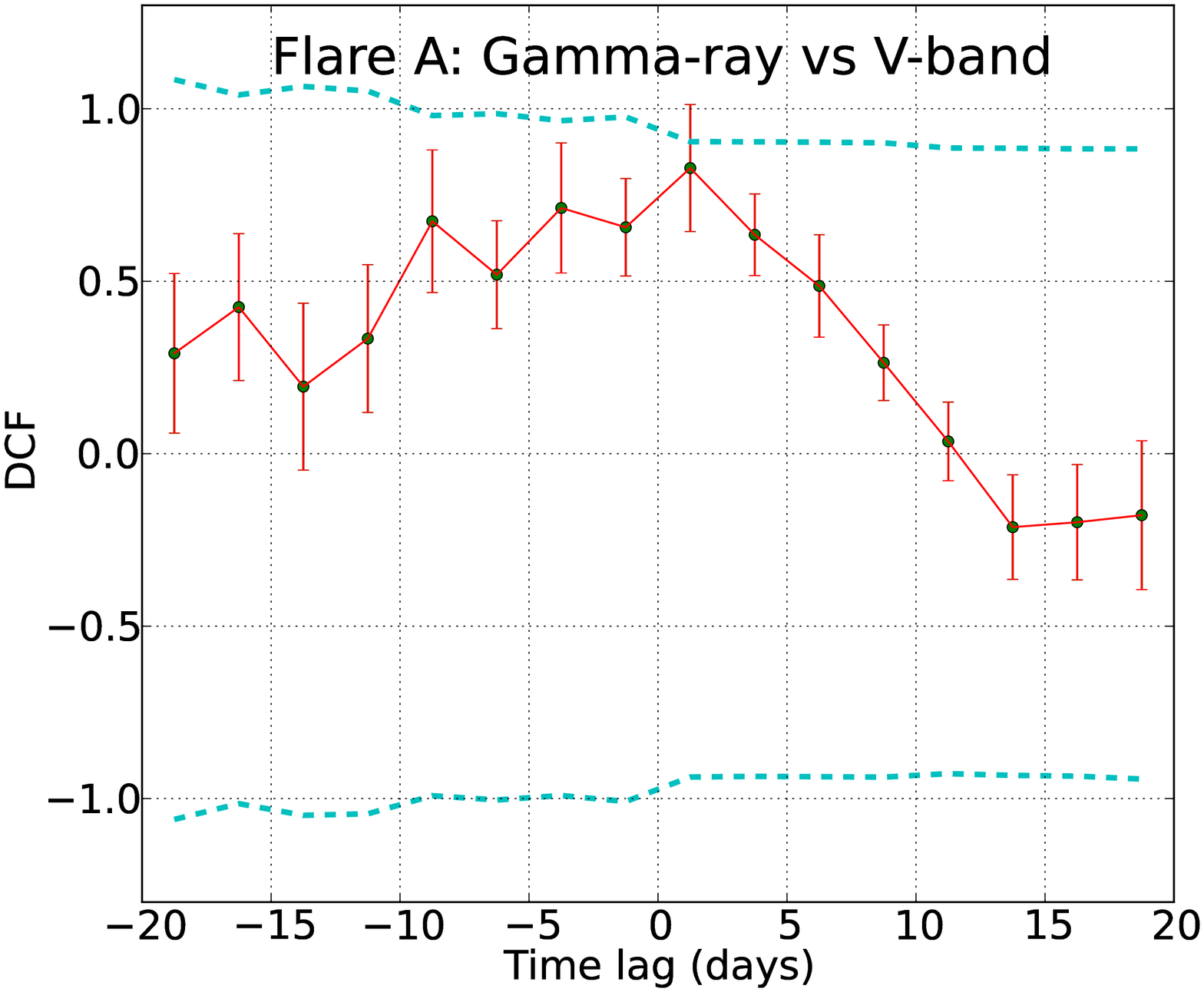}
 \includegraphics[scale=0.21]{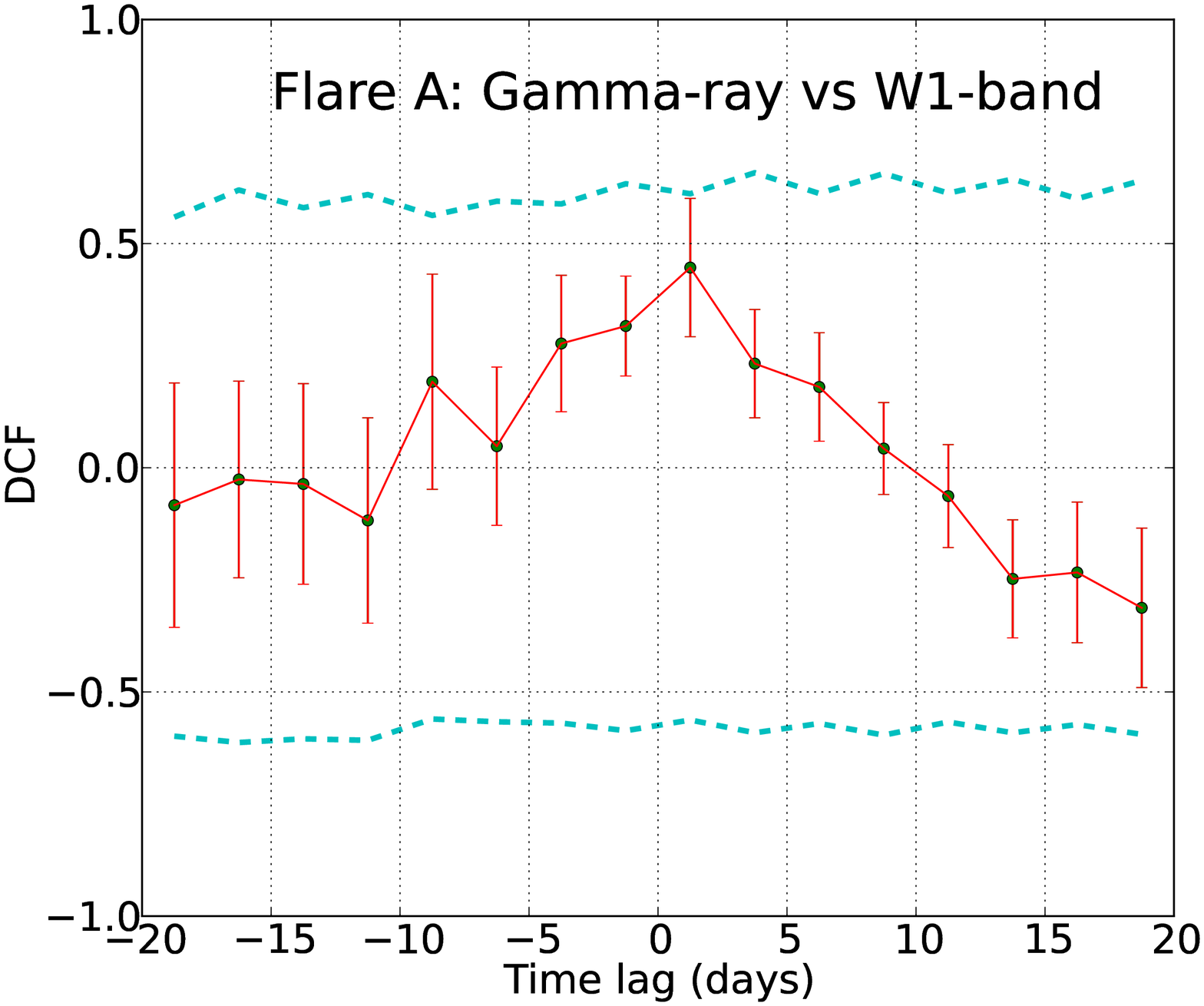}
 \includegraphics[scale=0.21]{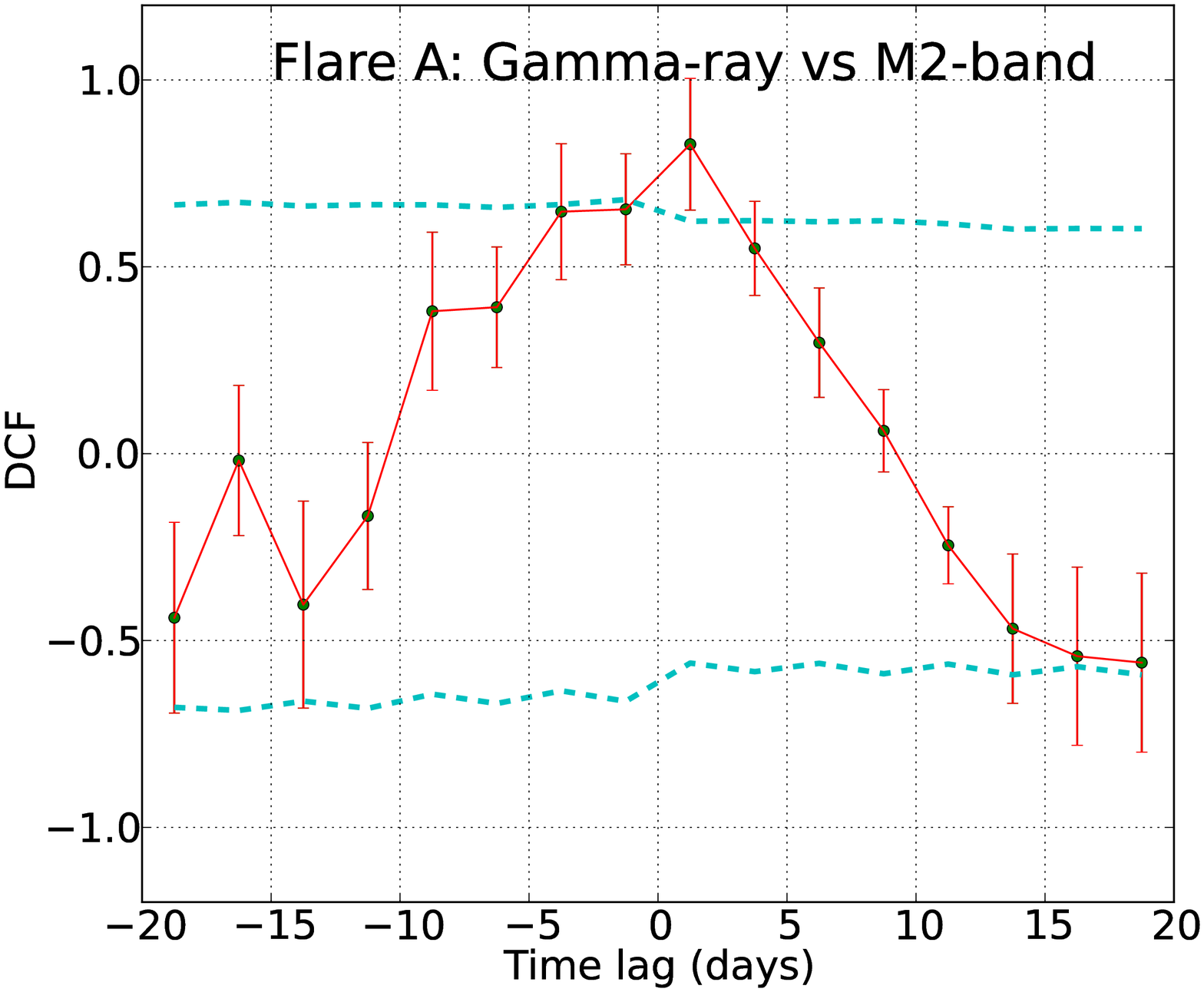}
 \includegraphics[scale=0.21]{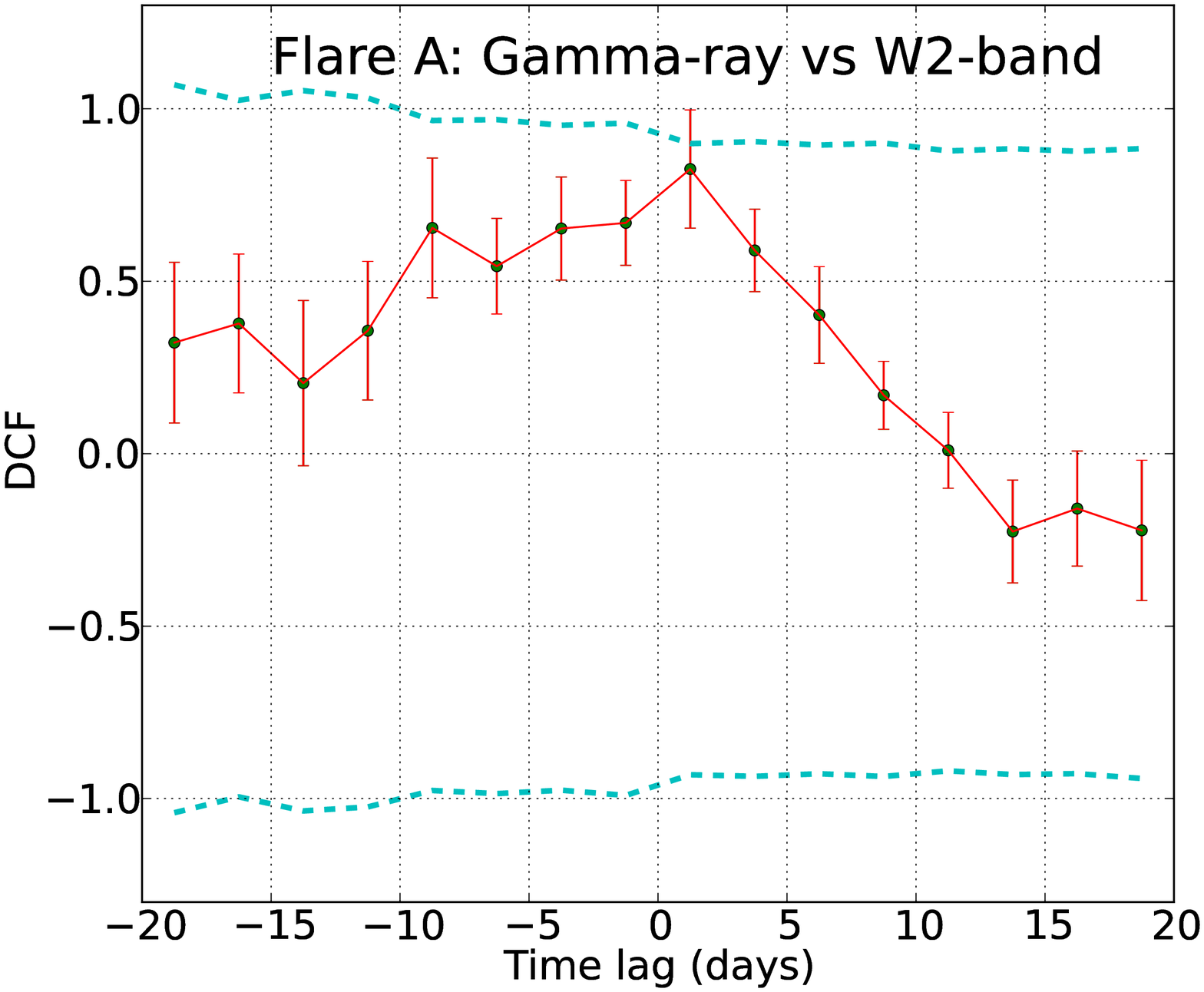}
 \caption{DCFs shown for ``Flare A'' for all the possible combinations: $\gamma$ vs. Swift-U, B, V, W1, M2, W2 band. The significance shown in
 cyan color is 95$\%$.}
\centering 
 \includegraphics[scale=0.21]{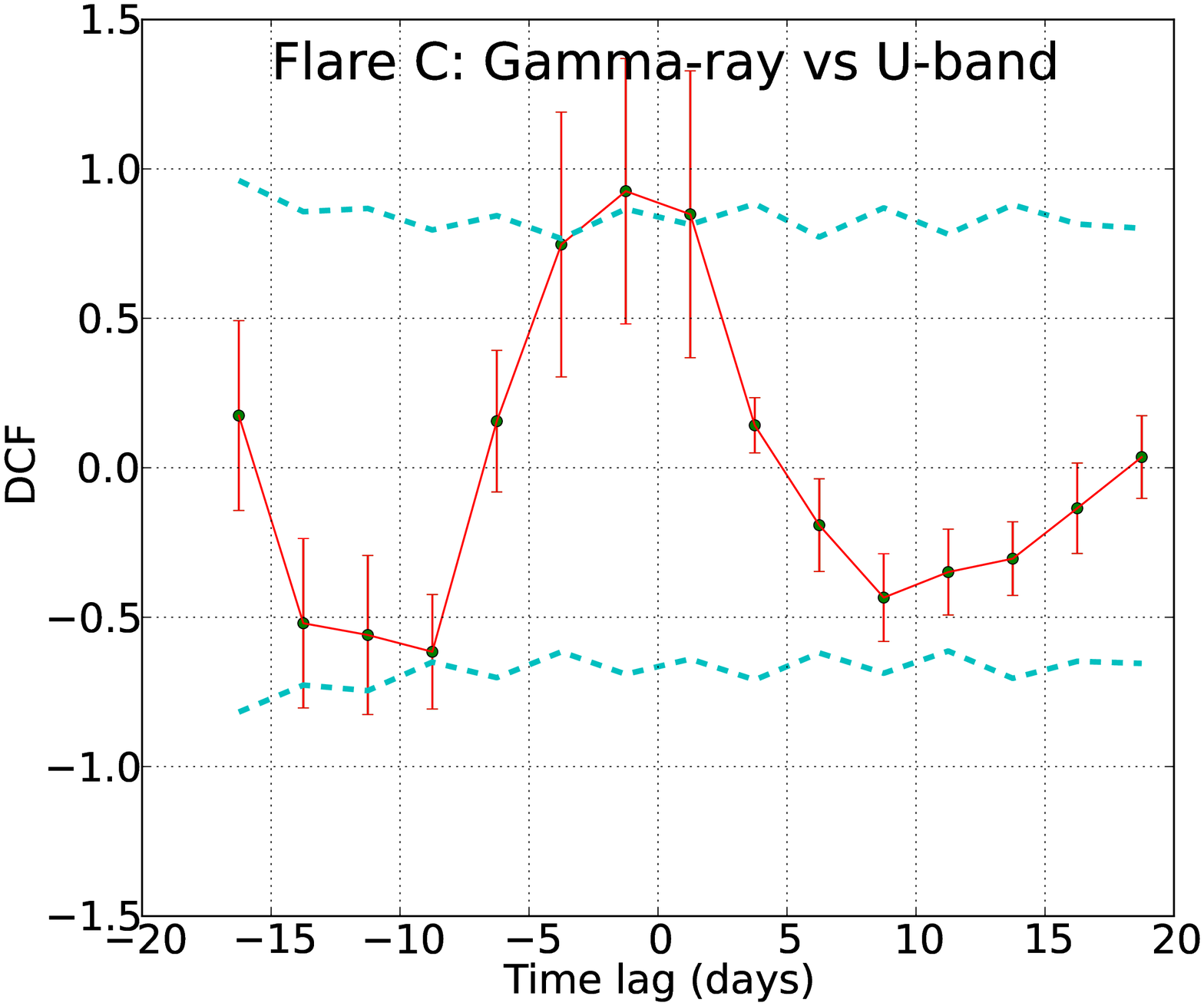} 
 \includegraphics[scale=0.21]{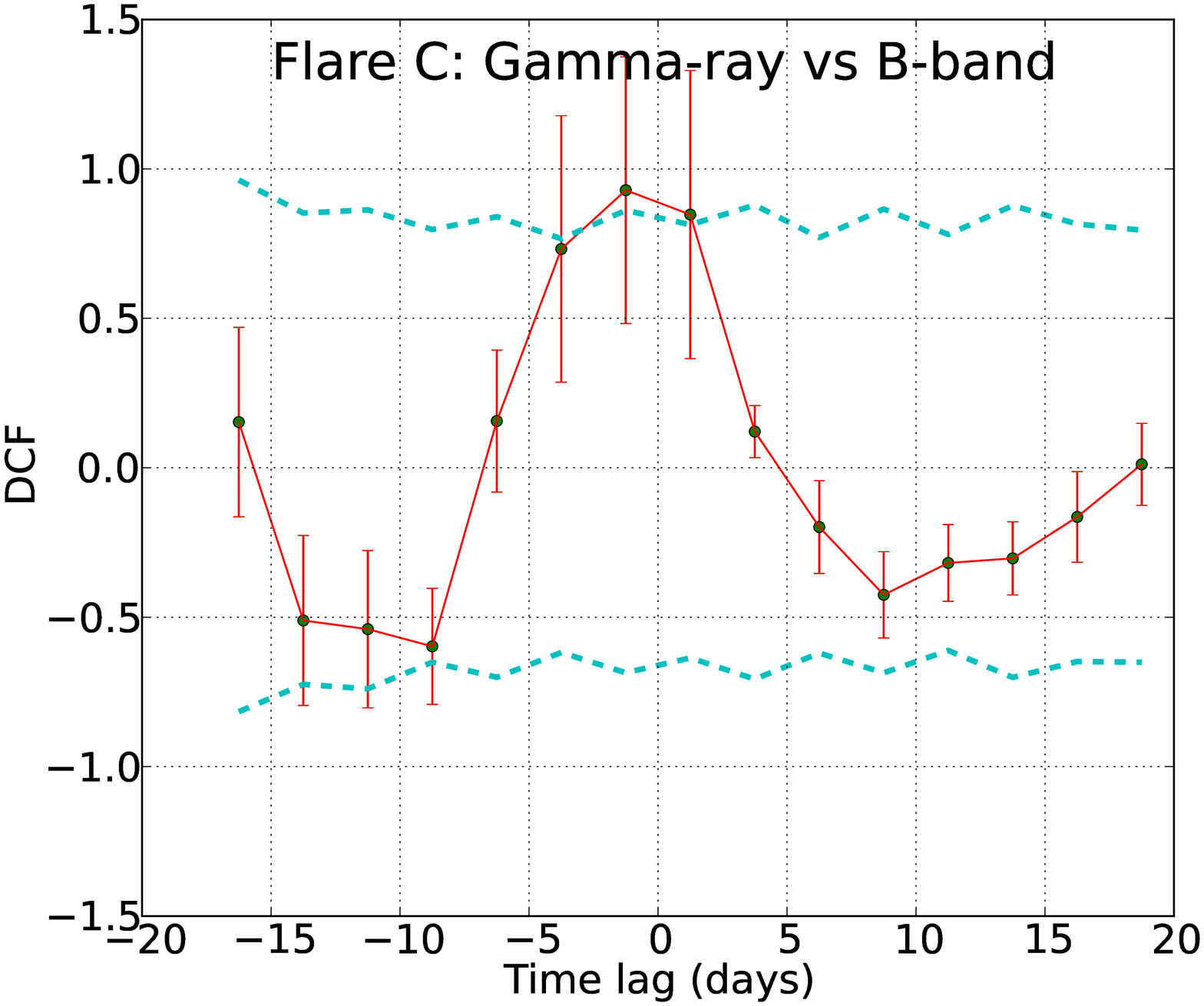}
 \includegraphics[scale=0.21]{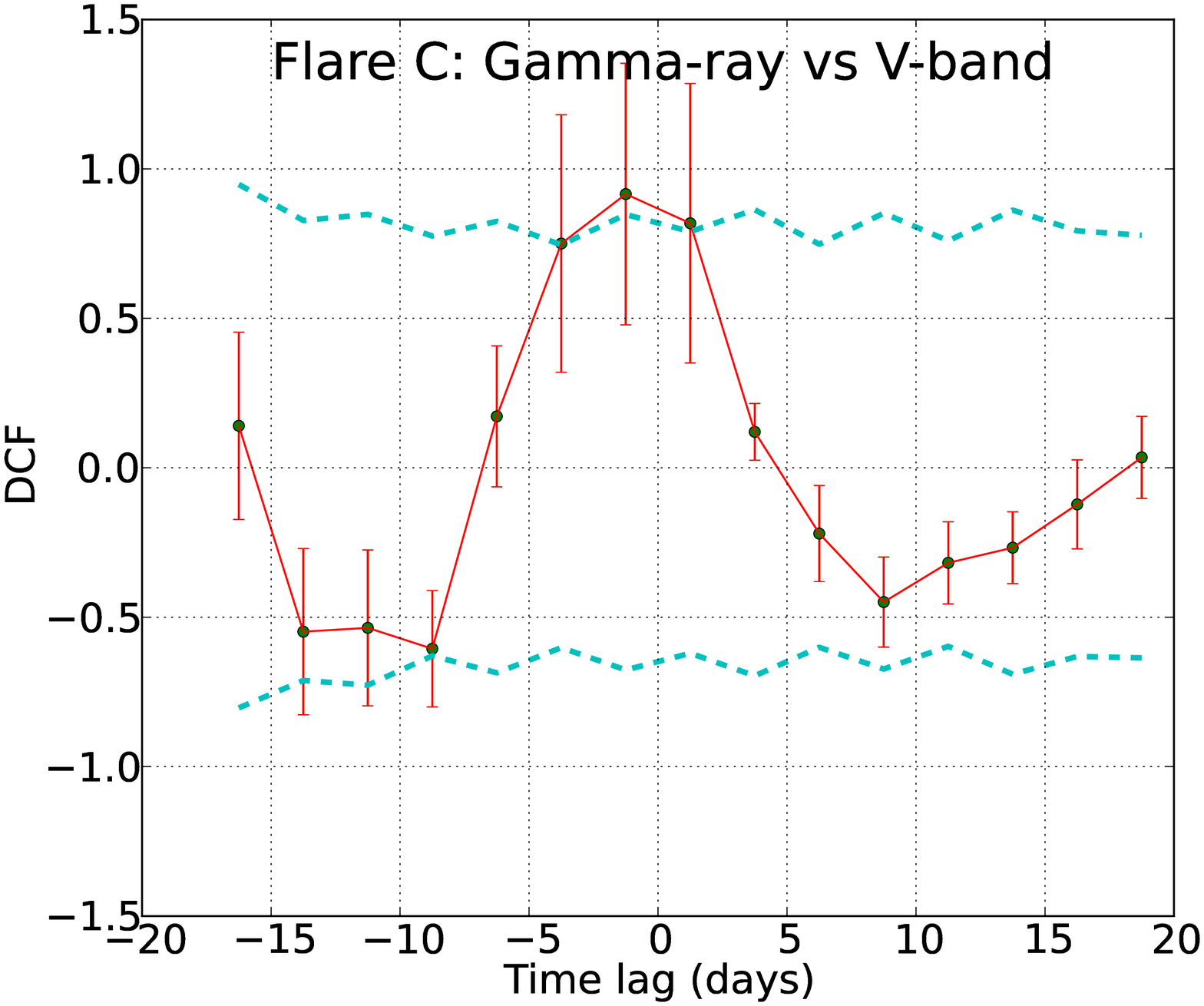}
 \includegraphics[scale=0.21]{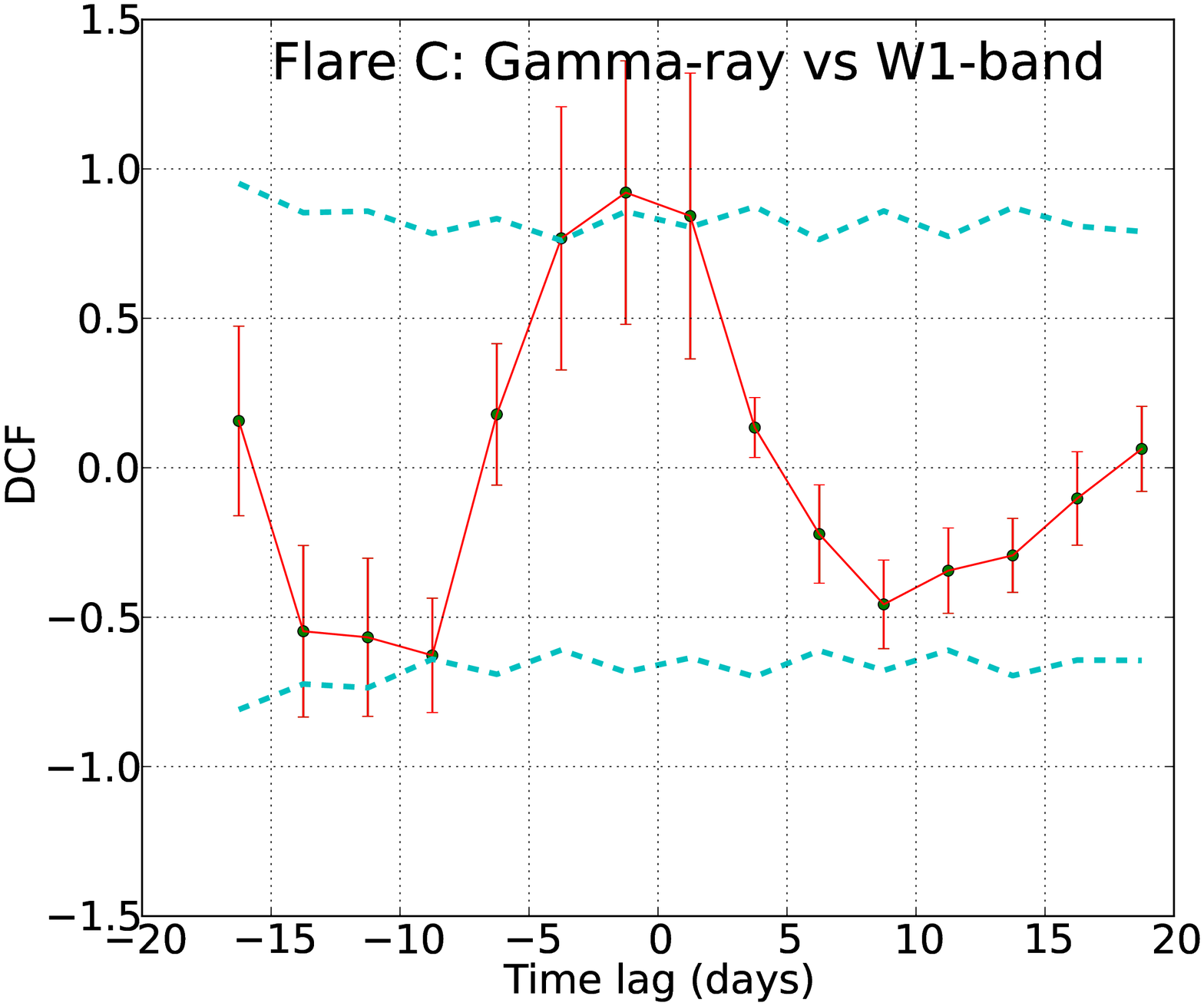}
 \includegraphics[scale=0.21]{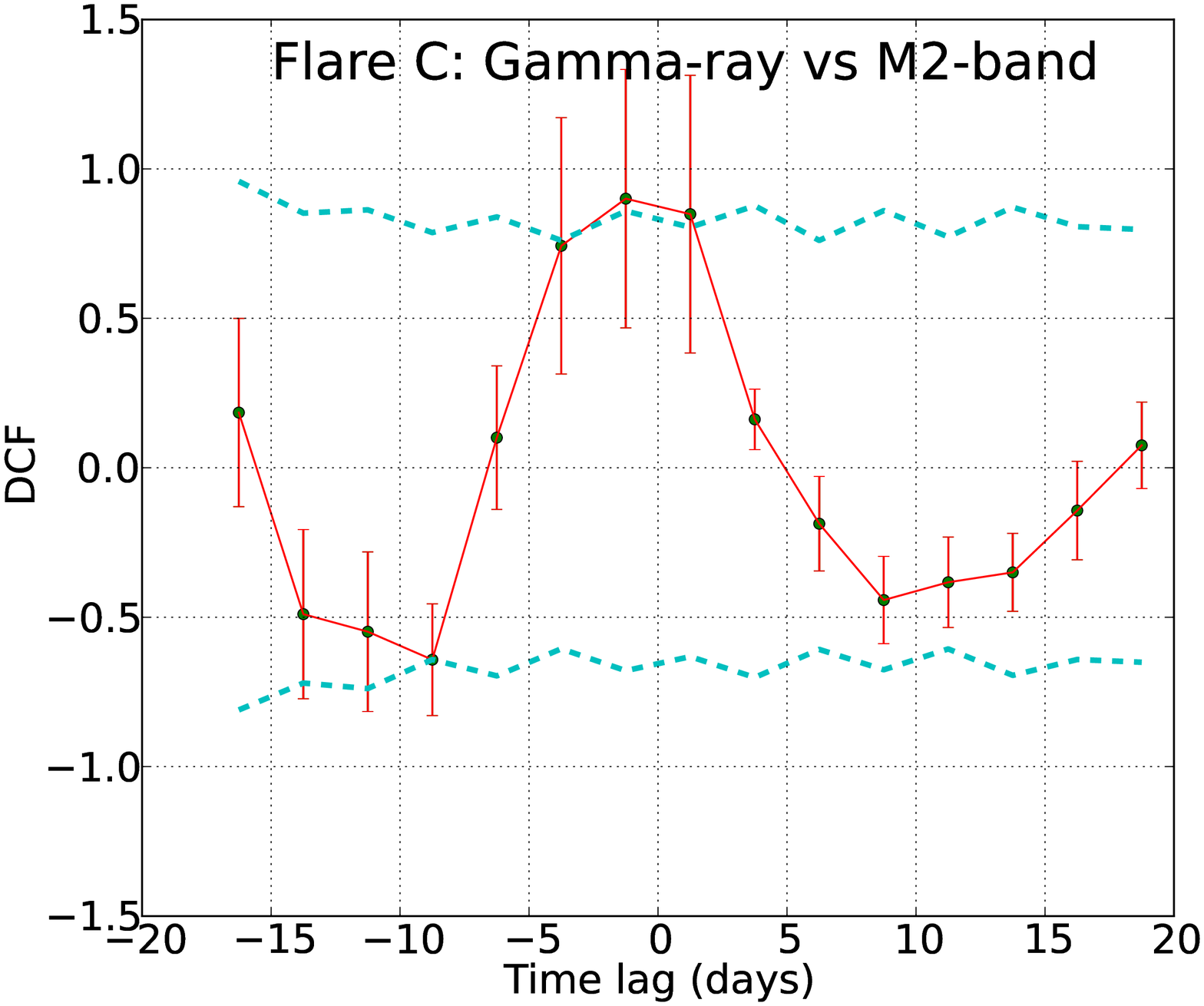}
 \includegraphics[scale=0.21]{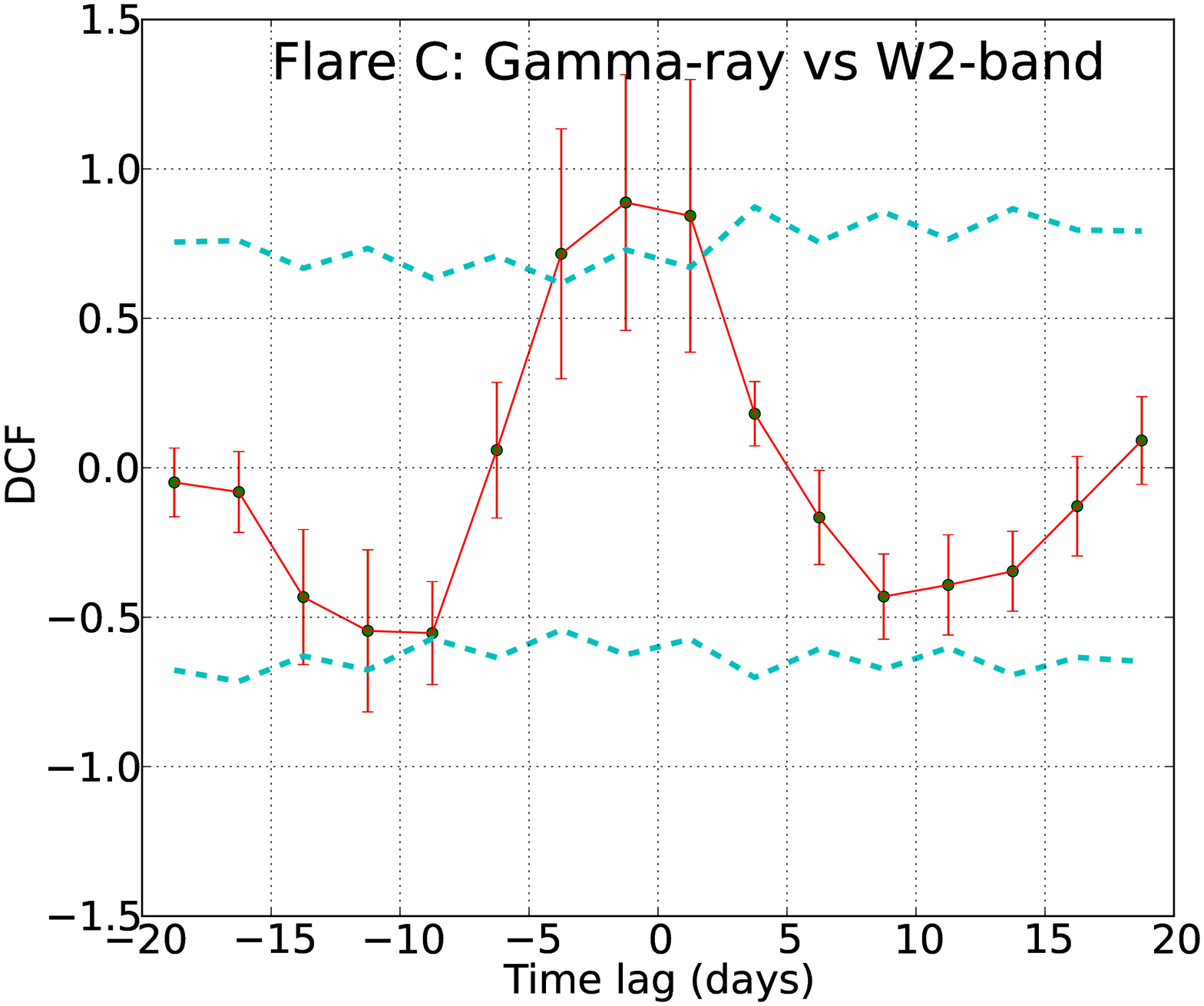}
 \caption{DCF shown for ``Flare C'' for all the possible combinations: $\gamma$ vs. Swift-U, B, V, W1, M2, W2 band. The significance shown in
 cyan color is 95$\%$.}
\vspace{-10pt}
\begin{center}
 \includegraphics[scale=0.20]{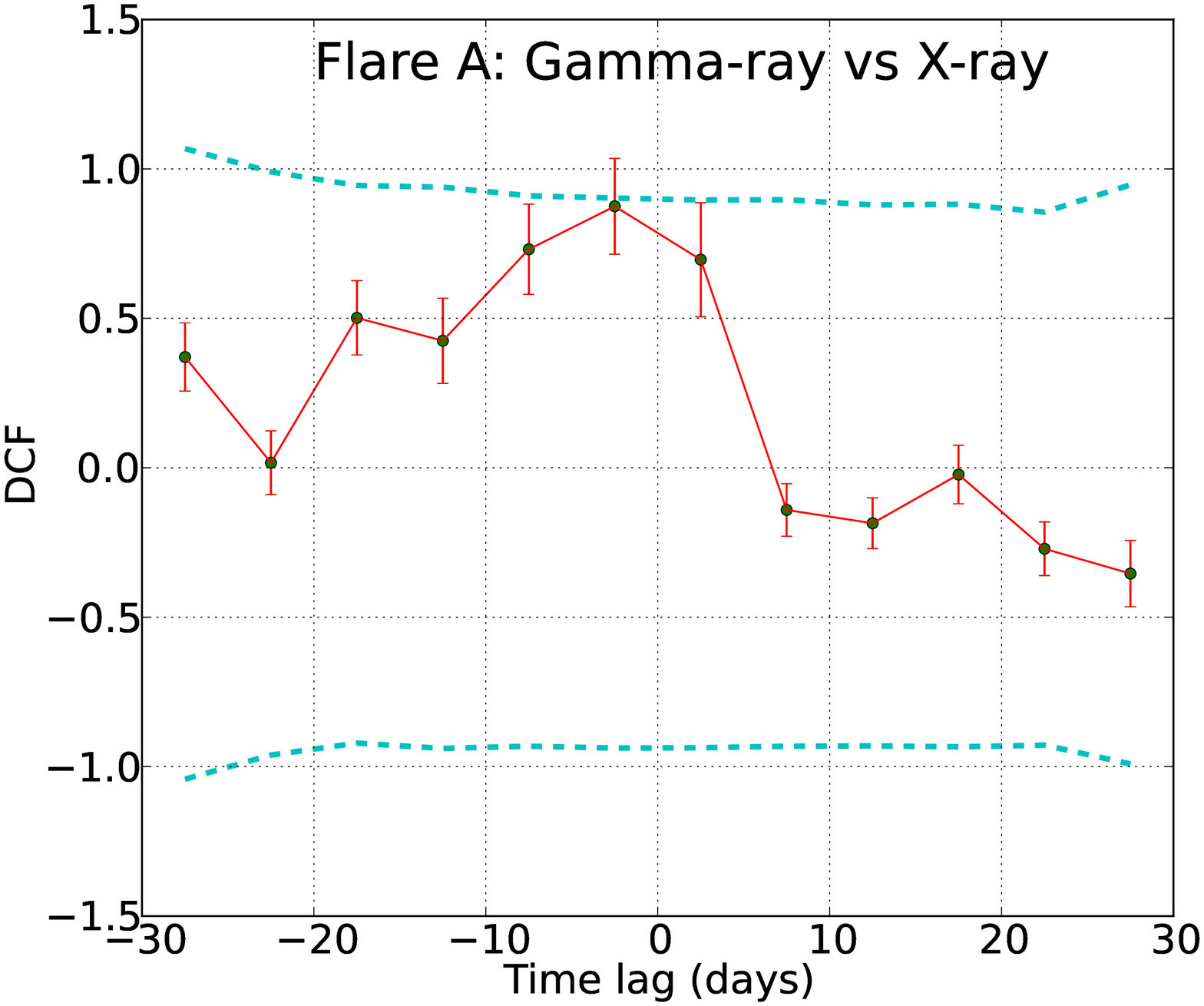}
 \includegraphics[scale=0.20]{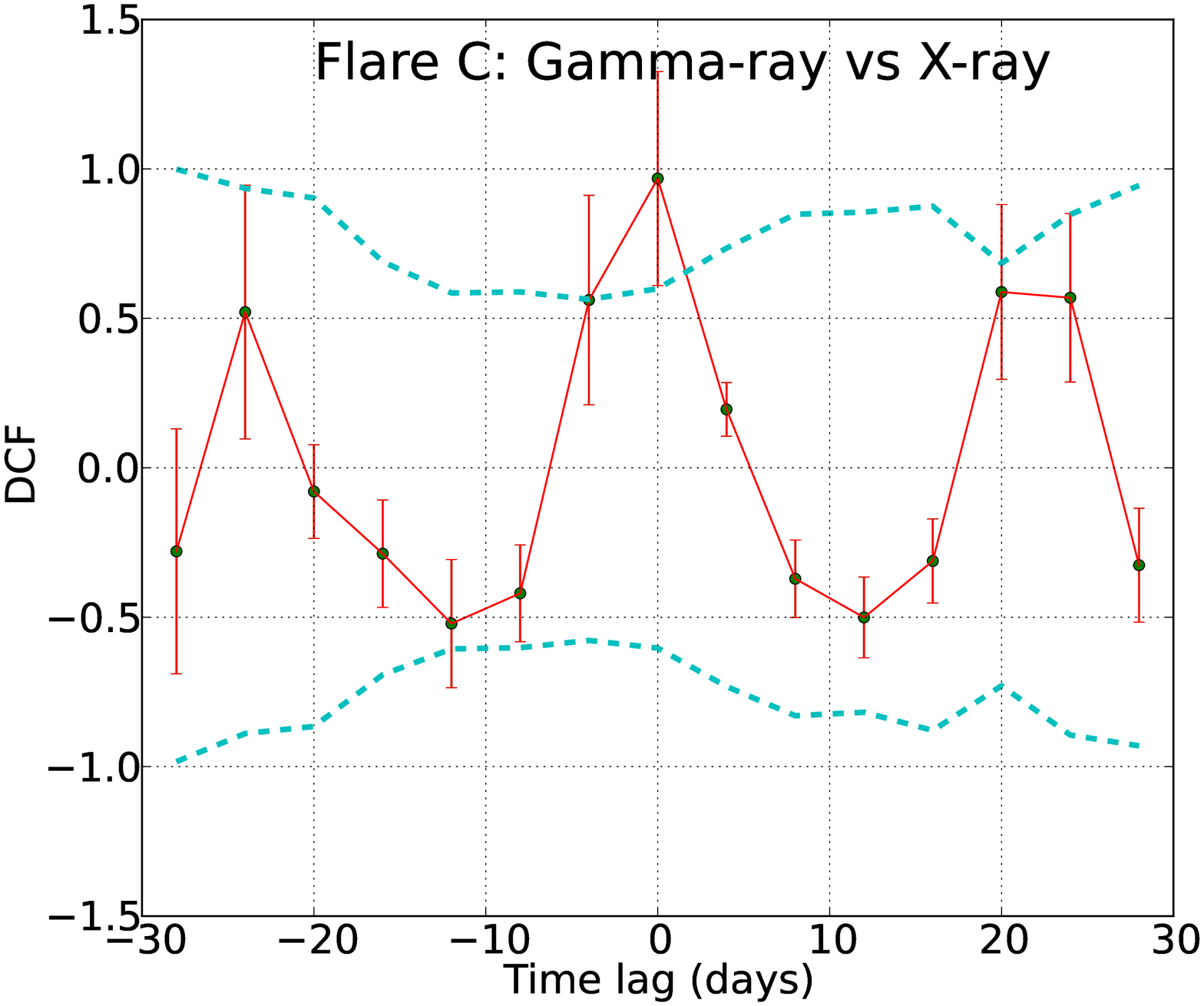}
 \end{center}
 \vspace{-15pt}
 \caption{Gamma-ray vs X-ray DCF are Shown for ``Flare A'' and ``Flare C''. The significance shown in cyan color is 95$\%$. }
\end{figure*}

%%%%%%%%%%%%%%%%%%%%%%%%%%%%%%%%%%%%%%%%%%%%%%%%%%%%%%%%%%%%%%

\section{Modeling the Multiwavelength SED}
The simultaneous observation of 3C 279 in different energy bands during its 2017--2018 flaring period provides an opportunity to gain further insight
into its multiwavelength properties.
In this work, theoretical modeling of the observed SED is done using the publicly available code GAMERA\footnote{http://joachimhahn.github.io/GAMERA}
\citep{Hahn (2015)} which solves the time-dependent transport equation and calculates the propagated electron distribution.

This electron spectrum finally used as an input to estimate the synchrotron, synchrotron self-Compton (SSC), and inverse Compton (IC)
emissions. The transport equation used in the code is following,
\begin{equation}\label{8}
\frac{\partial N(E,t)}{\partial t}=Q(E,t)-\frac{\partial}{\partial E}\Big(b(E,t) N(E,t)\Big)
\end{equation}
where, $Q(E,t)$ is the injected electron spectrum, and $N(E,t)$ is the spectrum achieve after the radiative loss. The radiative loss due to
synchrotron, SSC, and IC are represented by $b(E,t)$. GAMERA takes care of the inverse Compton in the Klein-Nishina regime from 
\citet{Blumenthal and Gould (1970)}. The transport equation does not have the diffusive loss term as it is insignificant compared to
the radiative loss of the electrons.
The log-parabola spectral model is chosen as the input injected electron spectrum, motivated from the gamma-ray spectral analysis.
To model the multiwavelength SEDs, my model considers a single spherical emitting zone or blob which is moving down the jet along the jet axis with a
Lorentz factor, $\Gamma$, and Doppler factor, $\delta$. The external photon field required for external Compton emission is believed to be dominated
by the broad line region (BLR) photons, particularly in FSRQ like 3C 279. The BLR photon density in the comoving frame is given by,
\begin{equation} \label{5}
 U'{_{BLR}} = \frac{\Gamma^{2} {\eta_{BLR}} L_{disk}} {4 \pi c R_{BLR}^{2}}
\end{equation}
where the $\eta_{BLR}$ represents the fraction of disk emission processed in BLR, I assume it to be typically 2\% \citep{Pittori et al. (2018)},
R$_{BLR}$ is the size of the BLR,
L$_{disk}$ denotes the disk luminosity, and c is the speed of light in vacuum. 
The photon energy density in BLR is only a fraction $\eta_{BLR}\sim 0.02$(2\%) of the disc photon energy density. %\citep{Ghisellini et al. (2009)}. 

The size of the BLR can be written as R$_{BLR}$ = 10$^{17}$L$_{d,45}^{1/2}$ (\citealt{Ghisellini and Tavecchio (2009)}), where L$_{d,45}$ is the 
accretion disk luminosity in units of 10$^{45}$ erg s$^{-1}$. For L$_{disk}$ = 2$\times$10$^{45}$ erg s$^{-1}$ \citep{Pian et al. (1999)}, the 
size of BLR is estimated to be R$_{BLR}$ = 1.414$\times$10$^{17}$ cm.

The minimum Doppler factor ($\delta_{min}$) during the flare can be estimated from the $\gamma\gamma$ opacity arguments and by estimating the highest
energy photon. The minimum Doppler factor can be calulated as (\citealt{Dondi and Ghisellini (1995)}; \citealt{Ackermann et al. (2010)}),
\begin{equation}
 \delta_{min} \cong \left[ \frac{\sigma_T d^2_L (1+z)^2 f_x \epsilon}{4 t_{var} m_e c^4} \right]^{1/6}
\end{equation}
which assumes that the optical depth of a photon ($\tau_{\gamma\gamma}$) with energy $\epsilon$ = $E$/m$_e$c$^2$ to the $\gamma\gamma$ interaction is 1. The luminosity
distance is denoted as d$_L$ (=3.1 Gpc), $\sigma_T$ is the Thompson scattering cross section, $E$ is the highest photon energy detected during the flare, 
$t_{var}$ is the variability time, and $f_x$ is the X-ray flux in 0.3-10 keV. Here, the values of $E$, $f_x$, and $t_{var}$ are estimated around the
same time period and the values are found to be 27 GeV at MJD 58228.45, 3.63$\times$10$^{-11}$ erg cm$^{-2}$ s$^{-1}$ at MJD 58228.38, and 1.14 days at
MJD 58228.50 respectively. The minimum Doppler factor is found to be $\delta_{min}$ = 10.7. The location of the gamma-ray emission region can be estimated by
assuming the bulk Lorentz factor $\Gamma$ = $\delta_{min}$ = 10.7, then the location can be defined as $d$ $\sim$ 2 $c\Gamma^2t_{var}$/(1+$z$).
It is found that the gamma-ray emission region is located at distance of 4.40$\times$10$^{17}$ cm from the central SMBH down the jet. This value is
comparable to the size of the BLR and hence conculded that during the emission of high energy photon (27 GeV) the gamma-ray emitting region must
have been located at the outer boundary of the BLR.

I have also considered the contribution of the accretion disk photons in the EC emission. The photon energy density in the comoving frame is defined
as (\citealt{Dermer and Menon (2009)}),
\begin{equation}\label{9}
U'_{disk}=\frac{0.207 R_g l_{Edd} L_{Edd}}{\pi c z^3 \Gamma^2}
\end{equation}
where, R$_g$ is known as gravitational radius, and l$_{Edd}$ = L$_{disk}$/L$_{Edd}$ is the Eddington ratio. Here z represents the distance of blob
from the SMBH and which is estimated to be 4.40$\times$10$^{17}$ cm. For black hole mass M$_{BH}$ = 2.51$\times$10$^{8}$ M$_{\odot}$ \citep{Wu et al. (2018)}, the
gravitaional radius found to be R$_g$ = 3.72$\times$10$^{13}$ cm. In this study, I have not considered dusty torus as a external target photon
field since there is no observational evidence. However, the contribution of NIR/optical/UV photons emitted by disk and dusty torus based clouds 
irradiated by a spine-sheath jet could be important (\citealt{Finke (2016)}; \citealt{Gaur et al. (2018)}; \citealt{Breiding et al. (2018)}) for 
the EC emission in some cases.

There are numbers of parameters that GAMERA uses as a input to model the multiwavelength emissions. The spectral index ($\alpha$, $\beta$), 
minimum and maximum ($\gamma_{min}$, $\gamma_{max}$) energies, magnetic field inside the blob (B),
and luminosity in injected electrons are the
parameters that have been optimized to obtain the best model fit to the SEDs. The BLR photon density ($U'{_{BLR}}$), BLR temperature = 10$^{4}$K
\citep{Peterson (2006)}, accretion disk photon density ($U'{_{disk}}$), disk temperature = 2.6$\times$10$^{6}$K \citep{Dermer and Menon (2009)}
 are kept fixed while modeling the SEDs.

The size of the blob can also be estimated by using the variability time and the minimum Doppler factor by
the relation,
\begin{equation}
 R \leq c t_{var} \delta (1+z)^{-1}
\end{equation}
where t$_{var}$ is the observed flux variability time (1.14 days). The size of the emitting blob is estimated to be 
R = 2.1$\times$10$^{16}$ cm. However, during the SED modeling Doppler factor and the size of the blob are optimized to best fit value.

A successful SED modeling is performed for all the three flares and one quiescent state. The best fit model parameters are shown in Table 5. The
multiwavelength SEDs modeled with GAMERA are presented in Figure 10. The low energy synchrotron peak is constrained by the optical and UV emission
and the high energy Compton peak by gamma-ray data points. The X-ray emission observed by Swift-XRT constrains the SSC peak. 
More magnetic field value is needed to explain the optical/UV emission in quiescent state, which suggest that the synchrotron process is more 
dominant here compared to the flaring state. The maximum electron energy found during all the states are in very much agreement though it is little
higher during the flaring state. 
The total jet power is estimated by the following relation,
\begin{equation}
 \rm{P_{jet}} = \rm{\pi}\  \Gamma^2 \ \rm{r^2}\  c \ (U_e+U_B+U_p)
\end{equation}
where $\Gamma$ is the Lorentz factor, $r$ is the size of the blob. The energy density in electrons, magnetic field, and cold protons are represented by
U$_e$, U$_B$, and U$_p$. Here the jet composition consists of an equal number of non-thermal electrons and cold protons. The calculated jet powers
in all components are shown in Table 5. The jet power is calculated for $\Gamma$=15.5 \citep{Jorstad et al. (2005)} and $r$ = 4.64$\times$10$^{16}$ cm.
It is found that the magnetic field has more jet power during the quiescent state compared to the flares.
However, the jet power required in electrons and protons is much higher in case of flaring state compared to the quiescent state. 
\\
The previous study on flares of 3C 279 suggest that most of the time the emission region is located outside the BLR (\citealt{Dermer et al. (2014)};
\citealt{Yan et al. (2015)}; \citealt{Vittorini et al. (2017)}). However, the brightest flare of 3C 279 during 2015 June reported by \citep{Paliya (2015a)},
demands a compact
emission region with high photon density and such regions can not be far from the central source. 
All earlier studies on this flare (June 2015) (\citealt{Hayashida et al. (2015)}; \citealt{Ackermann et al. (2016)}; \citealt{Pittori et al. (2018)})
found the emission region to be located within or at the boundary of the BLR.
Here, in this study too, similar result is obtained. A compact emission region can be inferred, which is located within the boundary of the BLR.

January 2018 flare of 3C 279
was also studied by \citet{Shah et al. (2019)}, where they have chosen BLR and IR photons as the target photons to describe the multiwavelength SEDs. They 
concluded that both cases can explain the broadband SEDs of various flaring states. However, the parameters found for EC/IR is more acceptable
than EC/BLR process. %My modeling shows that the SED can be explained by a single-zone model with BLR photons.
Two-zone leptonic and a single-zone lepto-hadronic SED modeling is performed by \citet{Paliya et al. (2015c)} to describe the broadband SEDs of 
3C 279 during the flare of 2013 December. However, my result shows that the single emission zone is enough to explain the broadband SEDs.

3C 279 is studied by \citet{Hayashida et al. (2012)} using the broadband data from radio to gamma-ray for the first two year (2008-2010) of 
\textit{Fermi} operation. They found a lag of 10 days between optical and gamma-ray emission during the flaring episode. However, they also
argue that the X-ray emission does not correlate with optical or gamma-ray emission. On the other hand my study exhibit a good correlation among
optical, X-ray and gamma-ray emission. They also have plotted the optical polarization (degree and angle) along with the gamma-ray flare and found huge
swing in the polarization angle corresponding to the gamma-ray flare (Period D). During the SED modeling, they constrained the location of the
emission region based on the observed change in the optical polarization angle. A huge swing in the polarization angle can be interpreted as
the precession in the jet, which allow the arbitrary location of the emission region within the BLR. My study also shows a huge variation
in the optical polarization angle during the flaring episode (Flare A and Flare B). Considering the polarization angle swing is caused by the
precession of jet, I have chosen the single emission zone within the boundary of the BLR, to model the broadband SED, which is also supported by 
the value estimated from the variability time.

Comparing the SEDs corresponding to quiescent and flaring state, I observed that the flux has increased during flares across the entire 
electromagnetic spectrum (Figure 10). The major flux change is observed in gamma-ray band between quiescent and flaring state. However, the change
in the gamma-ray flux among the different flares (A, B, \& C) are small. A relatively lesser flux change is observed in optical/UV and X-ray bands
among all the states. I have varied a few parameters to explain these changes and the parameters are magnetic field, minimum and maximum energy 
of electrons.
I found that almost ten times more jet power in electron is required to explain the gamma-ray flux observed during flares compared to the quiescent
state. 

\section{Summary}

\textit{Fermi}-LAT data was collected between 2017 November to 2018 July and three bright flares were observed. To differentiate among each other the
flares have been named as ``Flare A'', ``Flare B'', and ``Flare C''. A long low flux states period was observed just before the ``Flare A''. This low flux 
states is defined as quiescent state and time period of 50 days is chosen to represent it. 
A simultaneous observation in X-ray and Optical/UV was also done by 
Swift-XRT/UVOT telescope. The archival data from Steward observatory for optical V and R band has also been collected for the whole period, and a huge 
variation is seen in the degree of polarization and polarization angle. The archival radio data from OVRO and SMA are also collected for the entire flaring period.

The day scale variability has been seen in one-day bin light curve, which constrains the size of the emission region
to 2.1$\times$10$^{16}$ cm. The emission region is located at a distance of 4.40$\times$10$^{17}$ cm, which lies at the boundary of the BLR
(R$_{BLR}$=1.414$\times$10$^{17}$ cm). 
A ``harder-when-brighter'' trend were also observed in both gamma-ray and X-ray, which predicts detection of high energy photon during the high state.

The gamma-ray spectral data points are fitted with
three different spectral models PL, LP, and BPL. The TS$_{curve}$ obtained for all the fits suggest that LP is the best model to describe the 
gamma-ray photon spectrum. 
Further, I have estimated the fractional variability among different wavebands, and it is observed that the fractional variability is increasing 
towards the higher energy.

The discrete cross-correlation (DCF) study has been performed to find out the possible correlation in various wavebands emission.
The results show a good and strong correlation in all possible combinations like: gamma-ray vs. Optical/UV, gamma-ray vs. X-rays. In author's
knowledge this is the first time that a strong correlation with zero time lag between gamma-ray, X-rays, and optical/UV are seen for 3C 279.
A strong correlation and zero time lag between optical/UV, X-rays, and gamma-ray emissions suggest their co-spatial origin.
This outcome provides an impetus to choose single emission zone to explain the multiwavelength emission in SED modeling. 
\textit{GAMERA} is used to model the multiwavelength emission, and the parameters like magnetic field, injected electron spectrum, minimum and
maximum energy of injected electrons have been optimized to get a good fit to the SEDs data points. So this study suggests that a single-zone model
can also be good enough to explain the multi-waveband emissions from one of the brightest \textit{Fermi} blazar called 3C 279.

\begin{figure*} 
\centering
 \includegraphics[scale=0.38]{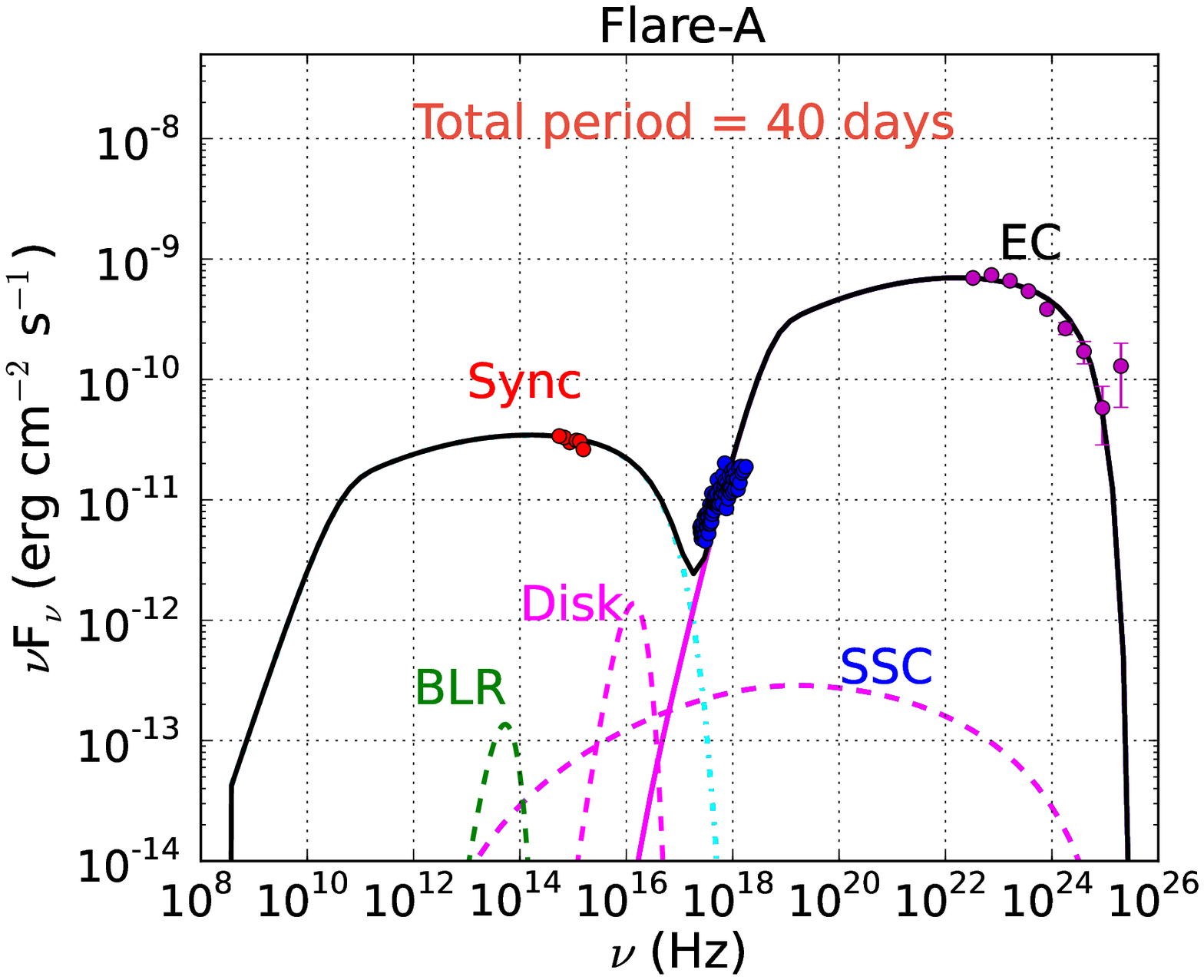}
 \includegraphics[scale=0.38]{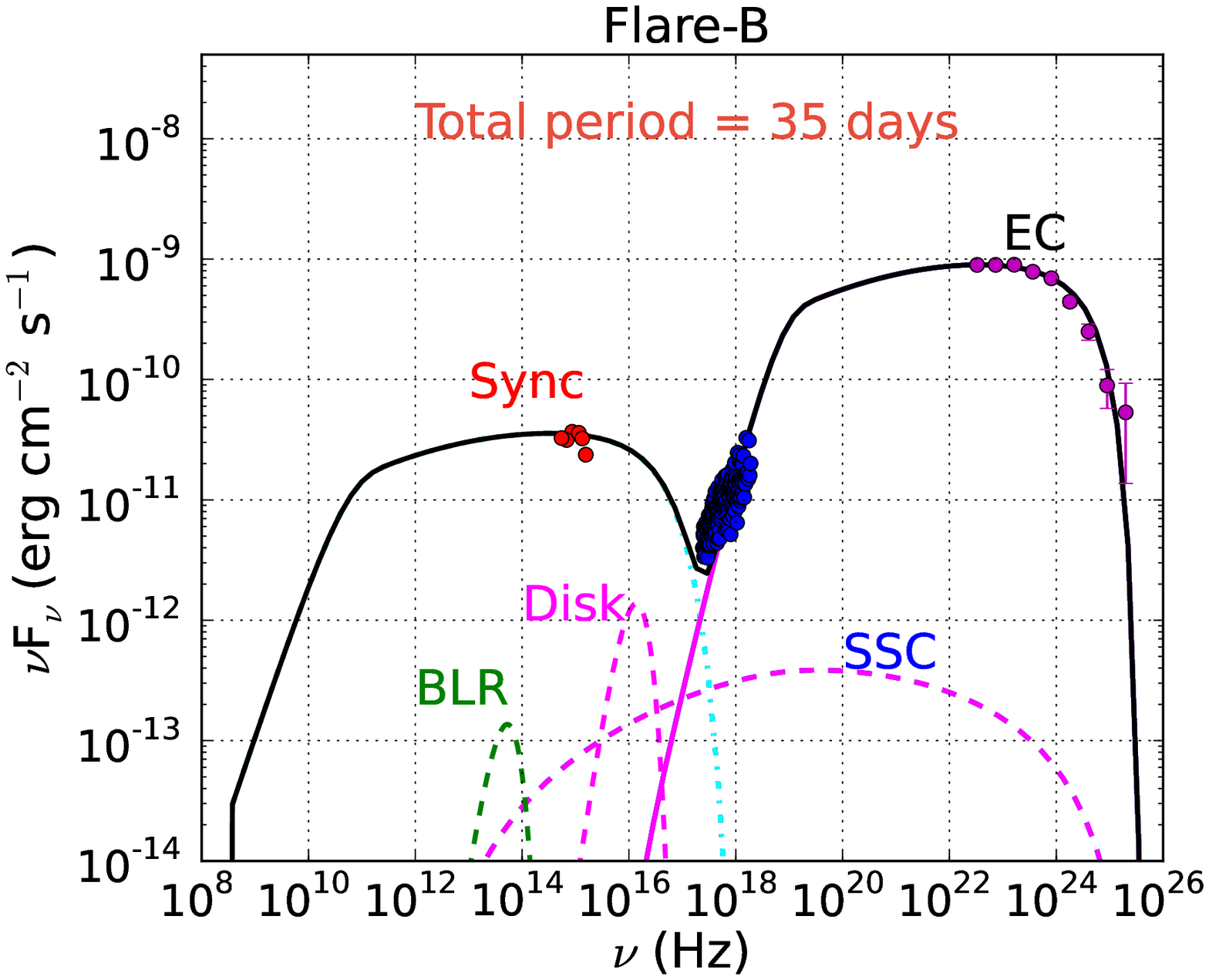}
 \centering
 \includegraphics[scale=0.38]{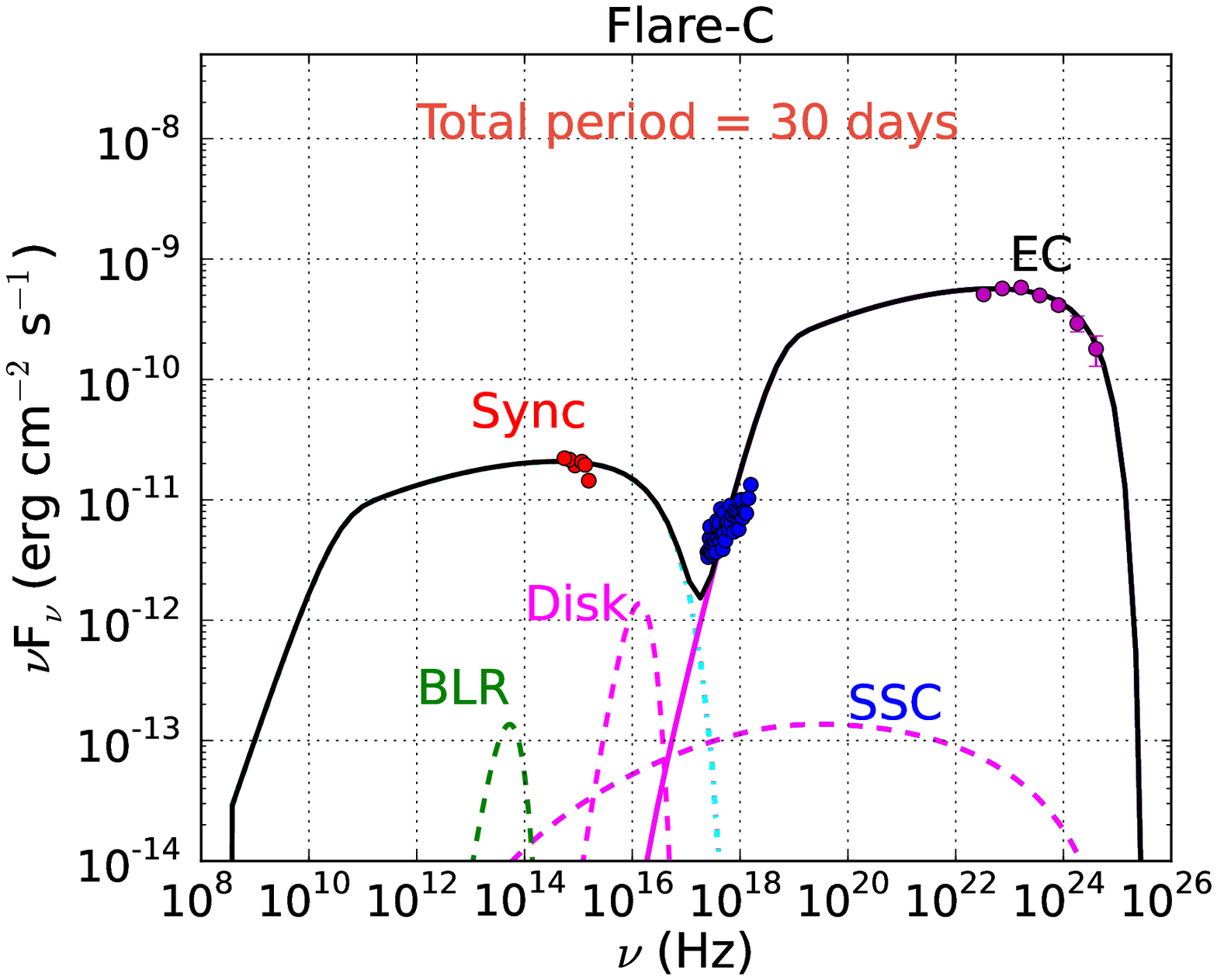}
 \includegraphics[scale=0.38]{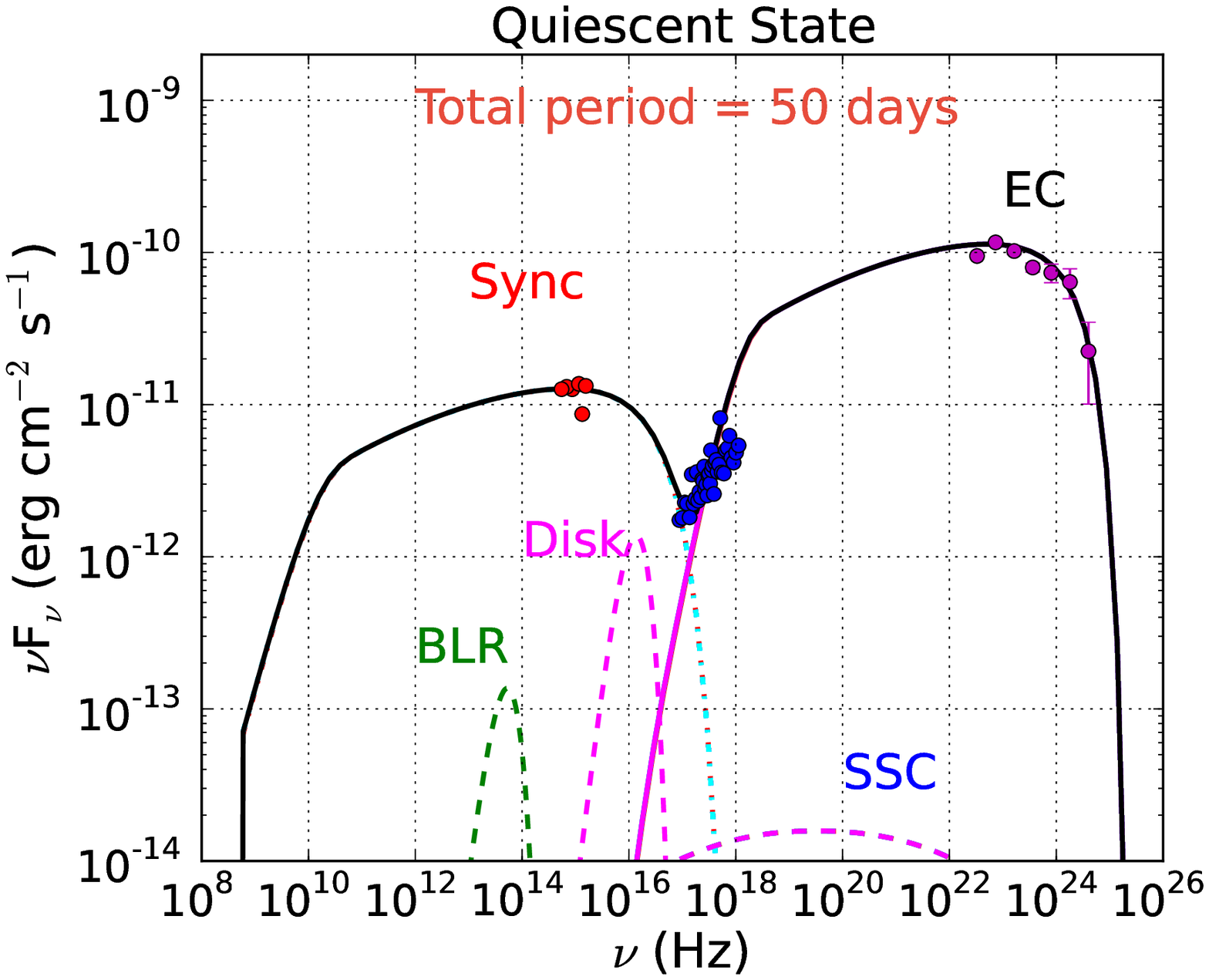}
 \caption{Multi-wavelength SEDs of all the flares and one quiescent state are presented. The optimize model parameters are shown in Table 5.}
\end{figure*}

\begin{table*}[htbp]
\begin{center}
\caption{ Results of fitted Multiwavelength SEDs showed in Figure 10. A LogParabola model is used as electron injected spectrum which is defined as
dN/dE = N${_0}$(E/E${_0}$)$^{(-\alpha-\beta*log(E/E{_0}))}$, where E${_0}$ is chosen as 90 MeV.}
\end{center}
\begin{tabular}{c c c c c}
\hline
 Activity &Parameters & Symbol & Values & Activity period (days) \\
\hline
 & BLR temperature & $T'_{blr}$ & 1$\times$10$^{4}$ K  \\
 & BLR photon density & $U'_{blr}$ & 6.377 erg/cm$^{3}$ \\
 & Disk temperature & $T'_{disk}$ & 2.6$\times$10$^{6}$ K  \\
 & Disk photon density & $U'_{disk}$ & 7.9$\times$10$^{-7}$ erg/cm$^{3}$ \\
 & Size of the emission region & R & 4.64$\times$ 10$^{16}$ cm   \\
 & Doppler factor of emission region& $\delta$ & 24.1 \\
 & Lorentz factor of emission region& $\Gamma$ & 15.5 \\
 \hline
  Flare A &&&  \\
 & Min Lorentz factor of injected electrons & $\gamma_{min}$ & 8.0  \\
 & Max Lorentz factor of injected electrons & $\gamma_{max}$ & 1.5$\times$10$^{4}$  \\
 & Spectral index of injected electron spectrum (LP) & $\alpha$ & 1.7 \\
 & Curvature parameter of LP electron spectrum & $\beta$ & 0.15 \\ 
 & magnetic field in emission region & B & 2.8 G & 40  \\
 & jet power in electrons & P$_{e}$ & 1.88$\times$ 10$^{45}$ erg/sec  \\
 & jet power in magnetic field & P$_{B}$ & 1.52$\times$ 10$^{46}$ erg/sec  \\
 & jet power in cold protons & P$_p$ & 1.87$\times$ 10$^{46}$ erg/sec \\
 \hline
 Flare B &&& \\
 & Min Lorentz factor of injected electrons & $\gamma_{min}$ & 10.0  \\
 & Max Lorentz factor of injected electrons & $\gamma_{max}$ & 1.7$\times$10$^{4}$  \\
 & Spectral index of injected electron spectrum (LP) & $\alpha$ & 1.7 \\
 & Curvature parameter of LP electron spectrum & $\beta$ & 0.12 \\
 & magnetic field in emission region & B & 2.5 G  & 35 \\
 & jet power in electrons & $P_{e}$ & 2.42$\times$ 10$^{45}$ erg/sec  \\
 & jet power in magnetic field & $P_{B}$ & 1.21$\times$ 10$^{46}$ erg/sec  \\
 & jet power in cold protons & P$_p$ & 2.24$\times$ 10$^{46}$ erg/sec \\
 \hline 
 Flare C &&& \\
 & Min Lorentz factor of injected electrons & $\gamma_{min}$ & 8.0  \\
 & Max Lorentz factor of injected electrons & $\gamma_{max}$ & 1.5$\times$10$^{4}$  \\
 & Spectral index of injected electron spectrum (LP) & $\alpha$ & 1.7 \\
 & Curvature parameter of LP electron spectrum & $\beta$ & 0.1 \\
 & magnetic field in emission region & B & 2.4 G & 30 \\
 & jet power in electrons & $P_{e}$ & 1.13$\times$ 10$^{45}$ erg/sec  \\
 & jet power in magnetic field & $P_{B}$ & 1.12$\times$ 10$^{46}$ erg/sec  \\
 & jet power in cold protons & P$_p$ & 1.56$\times$ 10$^{46}$ erg/sec \\
 \hline
 Quiescent state &&& \\
 & Min Lorentz factor of injected electrons & $\gamma_{min}$ & 4.0  \\
 & Max Lorentz factor of injected electrons & $\gamma_{max}$ & 1.2$\times$10$^{4}$  \\
 & Spectral index of injected electron spectrum (LP) & $\alpha$ & 1.7 \\
 & Curvature parameter of LP electron spectrum & $\beta$ & 0.08 \\
 & magnetic field in emission region & B & 4.2 G & 50 \\
 & jet power in electrons & $P_{e}$ & 2.56$\times$ 10$^{44}$ erg/sec  \\
 & jet power in magnetic field & $P_{B}$ & 3.42$\times$ 10$^{46}$ erg/sec  \\
 & jet power in cold protons & P$_p$ & 7.17$\times$ 10$^{45}$ erg/sec \\
\hline
\end{tabular}
\end{table*} 

\section*{Acknowledgements} \small

This work has made use of public \textit{Fermi} data obtained from FSSC.
This research has also made use of XRT data analysis software (XRTDAS) developed by ASI science data center, Italy. Archival data from the Steward
observatory is used in this research. This research has made use of radio data from OVRO 40-m monitoring programme \citep{Richards et al. (2011)}
which is supported in part by NASA grants NNX08AW31G, NNX11A043G, and NNX14AQ89G and NSF grants AST-0808050 and AST-1109911. The archival data 
from Submillimeter Array observatory has also been used in this study \citep{Gurwell et al. (2007)}. The Submillimeter Array is a joint project
between the Smithsonian Astrophysical Observatory and the Academia Sinica Institute of Astronomy and Astrophysics and is funded by the Smithsonian
Institution and the Acedemia Sinica. 
RP thanks Saikat, Manami, and Gunjan for manuscript reading and Arkadipta for helpful discussions regarding DCF.

\bibliographystyle{plain}

\end{document}